\documentclass[12pt, preprint]{aastex}
\usepackage{graphicx}
\def\gmin{\gamma_{\rm min}}
\def\gmax{\gamma_{\rm max}}
\def\gimin{\gamma_{\rm i, min}}
\def\gimax{\gamma_{\rm i, max}}

\def\fesy{f_{\epsilon}^{\rm sy}}
\def\feec{f_{\epsilon}^{\rm EC}}

\slugcomment{Submitted to ApJ}

\begin{document}

\title{Time-dependent Synchrotron and Compton Spectra from Jets of Microquasars}

\author{S. Gupta, M. B\"{o}ttcher}
\affil{Astrophysical Institute, Department of Physics and Astronomy,
Clippinger Hall 251B, Ohio University, Athens, OH 45701 -- 2979, USA}

\author{C. D. Dermer}
\affil{E. O. Hulburt Center for Space Research, Code 7653,
Naval Research Laboratory, Washington, D.C. 20375-5352, USA}

\begin{abstract}
Jet models for the high-energy emission of Galactic X-ray binary sources have regained 
significant interest with detailed spectral and timing studies of the X-ray emission 
from microquasars, the recent detection by the HESS collaboration of
very-high-energy $\gamma$-rays from the microquasar LS~5039, and the earlier
suggestion of jet models for ultraluminous X-ray sources observed in many 
nearby galaxies. Here we study the synchrotron and Compton signatures of time-dependent 
electron injection and acceleration, adiabatic and radiative cooling, and 
different jet geometries in the jets of Galactic microquasars.
Synchrotron, synchrotron-self-Compton, and external-Compton radiation 
processes with soft photons
provided by the companion star and the accretion disk are treated.
An analytical solution is presented to the electron kinetic equation 
for general power-law geometries 
of the jets for Compton scattering in the Thomson regime.
We pay particular attention to predictions concerning the rapid 
flux and spectral variability signatures expected in a variety of scenarios, 
making specific predictions concerning possible spectral hysteresis, similar
to what has been observed in several TeV blazars. Such predictions should be 
testable with dedicated monitoring observations of Galactic microquasars
and ultraluminous X-ray sources using Chandra and/or XMM-Newton.
\end{abstract}


\section{Introduction}

The high-energy emission from X-ray binaries (XRBs) is generally believed to be 
powered by the accretion of matter from a stellar companion onto a compact object
of typically a few solar masses. In many sources, this accretion process is
associated with the expulsion of collimated, mildly relativistic bipolar outflows 
(jets) most likely perpendicular to the accretion disk. In a few sources, e.g.,
GRS~1915+105 \citep{mr94}, VLBI radio images have shown a spatially-resolved radio jet
during episodes of quasi-steady and hard X-ray emission \citep{dhawan2000}.


In the standard picture, the high-energy (X-ray -- $\gamma$-ray) spectra of
X-ray binaries generally consist of two major components: A soft disk blackbody
with a typical temperature of $kT \sim 1$~keV, and a power-law at higher energies.
Neutron-star and black-hole X-ray binaries exhibit at least two main classes of
spectral states, generally referred to as the high/soft state, and the low/hard
state \citep[for a review see, e.g.][]{liang98,mcr04}. The high-energy spectra of X-ray 
binaries in the soft state are characterized by a thermal blackbody component, 
believed to be associated with thermal emission from an optically thick, 
geometrically thin accretion disk \citep{ss73}, and a power-law tail with a 
photon index $\Gamma\geq 2$. Generally, no high-energy cutoff of the high-energy 
power-law is detected. In the hard state, the spectrum is dominated by a power-law, 
with a slope $\Gamma < 2$ and a cut-off at $\sim$~a few hundred keV. 

While, in the conventional view of X-ray binaries (including microquasars), 
the X-ray and $\gamma$-ray emission is attributed to Comptonized emission 
\citep{st80,titarchuk94} arising from hot thermal ($kT\gg1 keV$) or relativistic, 
non-thermal electrons close to the black hole \citep[e.g.][]{lp77,bb77,sle76,ny94,
chen95}, the tentative {\it EGRET} detections of at least two Galactic 
microquasars at MeV --GeV $\gamma$-ray energies, namely LS 5039 \citep
{paredes00} and LSI+$61^\circ$303 \citep{gt78,taylor92,kniffen97}, the 
detection of X-ray jet structures in several microquasars using {\it Chandra} 
and {\it XMM-Newton} \citep[e.g.,]{corbel02, tomsick02}, and, most recently, 
the detection of very-high-energy (VHE) $\gamma$-ray emission from LS~5039 
\citep{aharonian05} have re-ignited interest in jet models for the high-energy 
emission from microquasars, analogous to the commonly favored models for 
blazars \citep[for a recent review see, e.g.] []{boettcher02}.

A jet origin of the X-ray emission of microquasars has previously been 
suggested by several authors, e.g., \cite{markoff01,markoff03a,markoff03b}, 
who discussed the possibility of synchrotron emission from relativistic 
electrons in the jet extending from the radio all the way into the X-ray 
regime. This idea is particularly well motivated by the tight temporal 
correlation between the radio and hard X-ray emission in microquasars, 
apparently anti-correlated with the soft X-ray emission 
\citep[e.g.][]{corbel00,corbel01}. 

Additional motivation for the investigation of jet models for the X-ray
emission of microquasars comes from the suggestion 
by \citet{georg02} that ultraluminous X-ray sources (ULXs, with X-ray luminosities 
$L_X \gtrsim 10^{39}$~erg~s$^{-1}$), detected in many nearby galaxies \citep[see, 
e.g.][]{makishima00,fabbiano01}, may be microquasars viewed at very small
angles with respect to the line of sight (hence dubbed ``microblazars''). 
However, more recent observational evidence, including indications of 
predominantly thermal spectra of ULXs \citep[e.g.][]{miller03} and the 
discovery of X-ray ionized nebulae around the optical counterparts of 
ULXs \citep{pm03,kaaret04}, provide increasing support for alternative 
models. 

Motivated by the new observational results summarized above, various
authors have been working on the expected high-energy emission signatures
from microquasars. In addition to the jet-synchrotron models mentioned
above, \cite{bosch05a,bosch05b} and \cite{romero05} have investigated 
various scenarios for the leptonic or hadronic origin of high-energy emission
from microquasars, aiming, in particular, at explaining the potential
{\it EGRET} detections of LS~5039 and LSI+$61^\circ$303. Those papers 
concentrate on a detailed description of the high-energy emission
processes in a steady-state configuration. However, if leptonic
emission mechanisms play a dominant role in the high-energy radiation
of microquasars, then one would expect to see significant spectral
variability, possibly similar to the spectral hysteresis observed
in the X-ray spectra of several high-frequency peaked BL~Lac objects,
such as Mrk~421 \citep{takahashi96} or PKS~2155-304 \citep{kataoka00}.

Indications of such behavior have been seen in XTE J1550-564 \citep{rct03}, 
but X-ray variability correlated with flaring behaviors detected at 
GeV energies with GLAST or at TeV energies with the ground-based air 
Cherenkov telescopes would provide strong evidence that leptonic 
processes are responsible for the $\gamma$-ray emission. A detailed 
time-dependent analysis of the high-energy emission signatures of 
jet models of microquasars has so far been restricted to a study of
the X-ray timing signatures of an empirical power-law model for the
high-energy emission \citep{kf04}, and a study of the expected QPO
features as well as broad-band spectral variability correlations on
longer time scales \citep{yuan05}.
 \cite{rkg03}\citep[see also][]{gkp04,g05} proposed a model of X-ray
variability from microquasar jets, in which the variability is driven by 
a variable soft photon source (the accretion disk), with a steady-state jet 
configuration. Those authors are focusing on predictions concerning the power 
density spectra as well as time and phase lag features, similar to previous 
work for accretion-disk corona models by \cite{kht97,hkt97,hkc98,bl98} and 
\cite{boettcher01}. No predictions concerning X-ray spectral hysteresis in HID
diagram form are made.

Since it is likely that many different radiation components (synchrotron,
synchrotron-self-Compton, external-Compton) are contributing to the 
high-energy emission from microquasars in the low-hard state, the 
rapid variability patterns expected in realistic jet models of microquasars 
might be more complex than the features previously investigated. In this paper, 
we present a detailed study of various plausible scenarios of electron injection
and acceleration into a relativistically moving emission region in a
microquasar jet, and subsequent adiabatic and radiative cooling. We
pay particular attention to the X-ray spectral variability, as motivated
above. In \S \ref{model}, we present a general outline of the assumed 
model geometry and choice of parameters for a baseline model used for 
our investigation. This section also contains a brief discussion of 
the emission mechanisms which we include in our calculations. In \S 
\ref{analytic}, we present an analytic solution to the electron 
kinetic equation in the case of power-law jet geometries, which 
is used for our time-dependent study of the radiation signatures from 
our model. Starting from our baseline model, we then perform a parameter
study to investigate the imprint of various parameter choices on the
expected spectral and variability patterns from microquasars, which 
is presented in \S \ref{parameterstudy}. We summarize in \S \ref{summary}.

\begin{figure}[t]
\includegraphics[height=10cm]{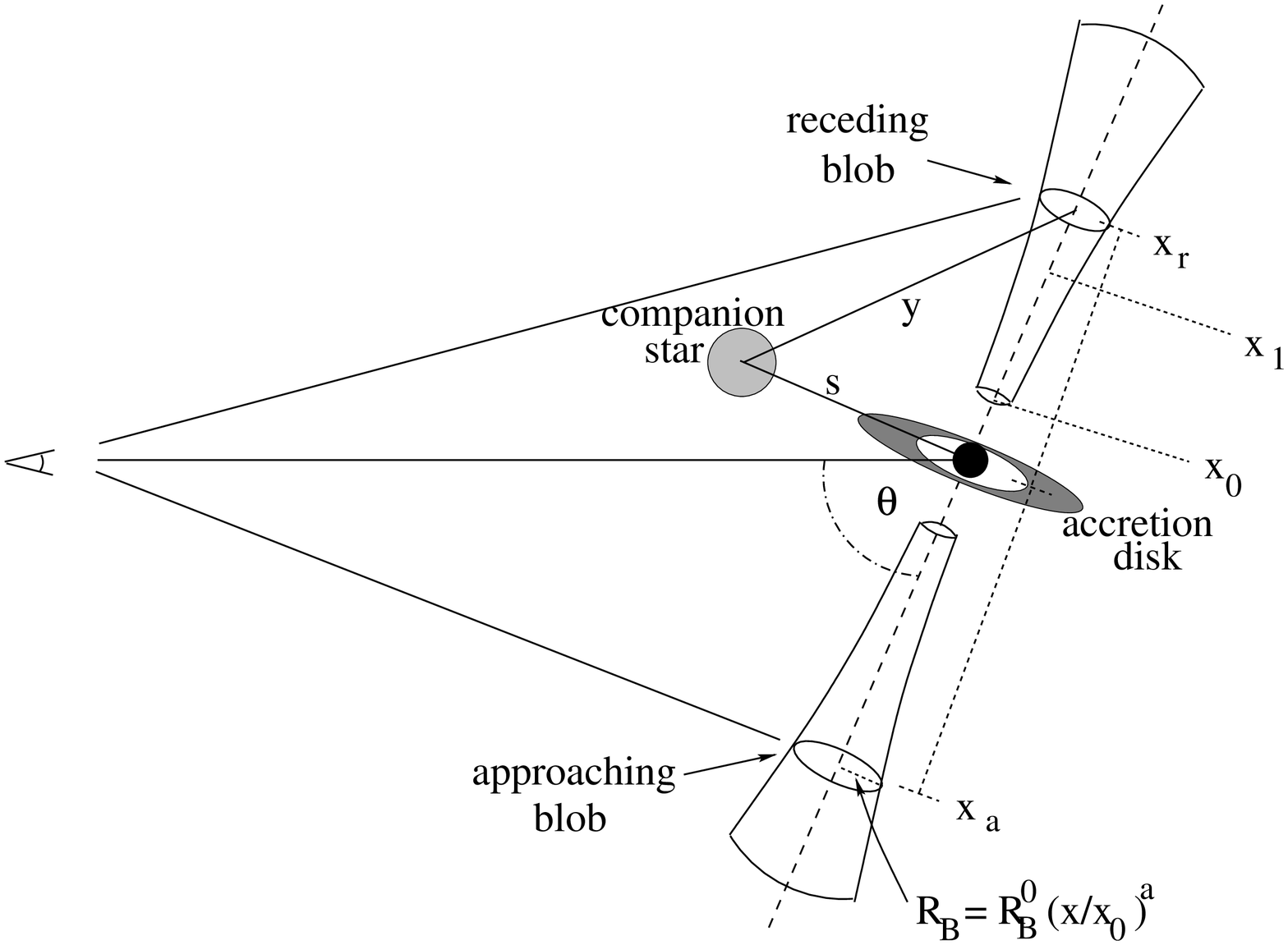}
\caption{Basic geometry of our model. See text for a detailed discussion.}
\label{geometry}
\end{figure}

\section{\label{model}Model Setup}

The geometry of our model setup is illustrated in Fig. \ref{geometry}.
The accretion flow onto the central compact object is ejecting a twin
pair of jets, directed at an angle $\theta$ with respect to the line of
sight. Two intrinsically identical disturbances, containing non-thermal
plasma (blobs) originate from the central source at the same time, 
traveling in opposite directions along the jet at a constant speed 
$v_j = \beta_j \, c$. Let $d$ be the distance to the source, and $\mu 
\equiv \cos\theta$. The time at which any radio component is observed,
is denoted by $t_{\rm obs}$. If we wish to distinguish between the 
approaching and the receding blobs, we do so using the subscript $a$ 
for the approaching and $r$ for the receding component of the jet. 
The intrinsic times (in the stationary frame of the central source) 
at which those components are being observed, are
\begin{equation}
t^{\ast}_{\rm a,r} = {t_{\rm obs} \over (1 \mp \beta_j \, \mu)},
\label{times}
\end{equation}
corresponding to linear distances of $x_{\rm a, r} = v_j \, 
t^{\ast}_{\rm a,r}$ from the central source. Here and throughout the
remainder of the paper, times $t^{\ast}$ measured in the rest frame of the central
source are denoted by a superscript `*', while times $t$ without
superscript refer to the co-moving frame of the blob. We assume that 
over a limited range in distance $x^{}_0 \le x^{} 
\le x^{}_1$, relativistic electrons are accelerated and injected
in the emission region. Most scenarios for the acceleration and injection of 
relativistic electrons in the jet will lead to a power-law spectrum of
injected electrons, which we parametrize in terms of an electron injection 
function,
\begin{equation}
{d^2 N_e (\gamma_i, t_i) \over d\gamma_i \, dt_i} = Q_0 \, \gamma_i^{-q}
\, H (\gamma_i; \gmin, \gmax) \; H (t_i; t_0, t_1),
\label{injection}
\end{equation}
where $H(y; y_0, y_1)$ is defined as 1 for $y_0 \le y \le y_1$ and 0
elsewhere, and $t_{0,1} = x_{0,1}/v_j$. The normalization constant 
$Q_0$ for the electron energy distribution is related to the injection
luminosity $L_{inj}$ through
\begin{equation}
Q_0 = \cases{
      \frac{L_{inj}(2-q)}{m_ec^2(\gamma_{max}^{2-q}-\gamma_{min}^{2-q})} & \mbox{if $q \not= 2$};
\cr
      \frac{L_{inj}}{m_ec^2 ln(\gamma_{max}/\gamma_{min})} & \mbox{if $q=2$}. 
\cr}
\end{equation}

The blob's (transverse) radius, $R_{\perp}$, scales with distance from the 
central engine as $R_{\perp} = R_{\perp}^0 \, (x / x_0)^{a}$, i.e., $a = 0$ 
corresponds to perfect collimation, and $a = 1$ describes a conical jet. In 
the following, we will consider values of  $0 \le a \le 1$, and for the
purpose of an analytical estimate, we assume no significant expansion 
along the direction of motion. Following the arguments given in
\cite{aa97}, we choose a magnetic-field dependence on distance from the
central black hole as $B (x) = B_0 \, (R_{\perp}/R_{\perp}^0)^{-2} = 
B_0 \, (x / x_0)^{-2a}$. 

Each electron injected into the emission region at relativistic
energies, will be subject to adiabatic and radiative cooling,
described by 
\begin{equation}
- {d\gamma \over dt} = {1 \over V_B} {dV_B \over dt} \, {\gamma \over 3} 
+ {4 \over 3} \, c \, \sigma_T \, {u \over m_e c^2} \, \gamma^2
\label{cooling1}
\end{equation}
in the co-moving frame of the blob, where the first term on the 
r.h.s. describes the adiabatic losses. The second term 
describes synchrotron and Compton losses, 
with $u = u_B + u_{\rm rad}$, where $u_B = B^2/ 8\pi$ is the magnetic-field 
energy density, and $u_{rad}$ is the seed photon energy density for 
Compton scattering in the Thomson regime. The term $u_{\rm rad}$ consists 
of contributions from the X-ray emission from an optically thick accretion disk,
from the intrinsic synchrotron radiation, and from external photons
from the companion star. From the milli-arcsecond resolution of the 
VLBA monitoring observations which have resolved superluminal-motion 
components in Galactic microquasars, we can estimate that such components 
appear on linear scales of $x^{} \sim 10^{-3} \, x^{}_{-3}$~pc 
with $x^{}_{-3} \sim 1$ -- 100. The energy density of disk photons 
in a point-source approximation is then $u_{\rm ext} \sim 3 \times 
10^{-5} \, L_{38} \, (x^{}_{-3})^{-2} \, \Gamma_j^{-2}$~ergs~cm$^{-3}$. 
The magnetic-field energy density of $u_B \approx 4 \times 10^{-2} \, 
B_{\rm G}^2$~ergs~cm$^{-3}$ where $B_{\rm G}$ is the magnetic field in Gauss. 
Given the estimated magnetic-field values of $\sim$ a few tenths of Gauss 
estimated on scales of several mpc for GRS~1915+105 \citep{aa99}, the 
magnetic-field decay mentioned above suggests initial magnetic field 
values of $B_0 \sim $ several thousand Gauss at injection sites $x_0^{}$ 
of a few thousand $R_g$ from the black hole. Since there does not seem 
to be any evidence for a ``Compton catastrophe'' in Galactic jet sources, 
we may assume that the soft photon energy density of the synchrotron 
radiation field is significantly lower than the magnetic-field energy 
density. However, in high-mass X-ray binaries (HMXB) like Cyg~X-1, 
LS~5039, or LS~I~+61$^o$303, it is possible that the energy density 
in the soft photon field from the companion star is dominant over 
the magnetic-field energy density. The soft-photon energy density 
from a companion star of luminosity $L_{\ast} = 10^{39} \, L_{39}$~ergs
~s$^{-1}$ at a separation of $s^{} = 10^{12} \, s^{}_{12}$~cm from the 
black-hole primary is $u_{\ast} \approx 3 \times 10^3 \, (\Gamma_j^2 \, L_{39} /
y^{2}_{12})$~ergs~cm$^{-3}$, where $y^{}_{12} = \sqrt{x^{2}_{12} + s^{2}_{12}}$ 
is the distance between the blob and the star in units of $10^{12}$~cm.

\begin{deluxetable}{lccc}
\tabletypesize{\scriptsize}
\tablecaption{Parameter choices for our baseline model}
\tablewidth{0pt}
\tablehead{
\colhead{Parameter} & \colhead{Symbol} & \colhead{Value}
}
\startdata
Black-hole mass                                 & $M$                & $15 \, M_{\odot}$ \\
Distance                                        & $d$                & $3.75 \times 10^{22}$~cm \\
Jet inclination angle                           & $\theta_{\rm jet}$ & $70^o$ \\
Bulk Lorentz factor                             & $\Gamma_j$         & $2.5$ \\
Binary separation                               & $s^{\ast}$         & $10^{12}$~cm \\
Luminosity of companion star:                   & $L_{\ast}$         & $8\times 10^{37}$~ergs~s$^{-1}$ \\
Surface temperature of the companion star       & $T_{\ast}$         & $3 \times 10^4$~K \\
Initial blob radius                             & $R_0$              & $10^3 \, R_g$ \\
Jet collimation parameter                       & $a$                & $0.3$ \\
Accretion Fraction                              & $\dot{m} = \dot{M}/\dot{M}_{\rm Edd}$ & $0.01$ \\
Accretion disk luminosity                       & $L_D$              & $1.9 \times 10^{37}$~ergs~s$^{-1}$ \\    
Electron injection spectrum, low-energy cutoff  & $\gamma_{\rm min}$ & 10 \\
Electron injection spectrum, high-energy cutoff & $\gamma_{\rm max}$ & $10^4$ \\
Electron injection spectrum, spectral index     & $q$                & 2.4 \\
Beginning of electron injection zone            & $x^{}_0$       & $10^3 \, R_g$ \\
End of electron injection zone                  & $x^{}_1$       & $10^5 \, R_g$ \\
Magnetic field at $x^{}_0$                  & $B_0$              & $5 \times 10^3$~G \\
Injection luminosity                            & $L_{\rm inj}$      & $4.4 \times 10^{-5}~L_{Edd}$ \\
\enddata
\label{par_table}
\end{deluxetable}

The standard parameter choices for our ``baseline'' model, are broadly representative 
of GRS 1915+105 in the low/hard state, which also give  equipartition between the energy densities of the 
relativistic electrons and the magnetic field in the ejecta. 
Repeated VLA observations of ejection events from radio emitting clouds of GRS 1915+105 
have established that the ejecta move with a true speed of $\beta = 0.92$, giving a bulk 
Lorentz factor of $\Gamma_j = 2.5$, at an angle $\theta=70^o$ with respect to the line of sight
\citep{mr94}.
The recent measurement of rotational broadening of the early-type K giant companion star,
combined with the orbital parameters lead to a black hole mass of $14\pm 4 \rm M_{\odot}$ 
\citep{hg04}. For our parameter study, we have adopted a black hole mass of $15 \rm M_{\odot}$.
Radial systemic velocity and radio data calculations \citep{fender99,greiner01} have 
yielded a distance of $d\sim 12 \rm kpc$ to the source.
Our reference values for the orbital separation of the binary and the donor star's luminosity 
follow from a mass function of $9.5\pm 3 \rm M_{\odot}$ derived by \cite{greiner01} from infrared 
observations.
For the disk, a mass accretion rate of $1\%$ in Eddington units, translating to $\sim$ a few times
$10^{-9} \rm M_{\odot}/yr$  for a $15 \rm M_{\odot}$ black hole, is a reasonable estimate in the low-hard 
state \citep{mcr04,bm00}.
We inject electrons into the jet over a distance of about 100$R_g$, and follow the evolution
of the jet for about $5$ days.
The magnetic field values in the jet are hard to constrain, but we start at an initial value of
$B_0 = 5000 \rm G$ at the base of the jet, resulting in $B\sim 0.2-0.3 \rm G$ after $\sim 4.5$ days from
the start of ejection \citep{aa99}. The above parameter choices are summarized in 
Table ~\ref{par_table}.


For the purpose of an analytic treatment of the electron kinetic
equation, Eq. \ref{cooling1}, we re-write it in the form
\begin{equation}
- {d\gamma \over dt} = m \, {\gamma \over t} + \left( \nu_0 \, 
\left[{t \over t_0} \right]^{-4 a} + \nu_{\rm D} \, \left[{t \over t_0} \right]^{-2}
+ f_{\ast} (t) \right) \, \gamma^2
\label{cooling2}
\end{equation}
with $m = 2a/3$,
\begin{equation}
\nu_0 \equiv {4 \over 3} \, c \, \sigma_T \, {B_0^2 \over 8 \pi \, m_e c^2}
\label{nu_0},
\end{equation}
\begin{equation}
\nu_{\rm D} \equiv {4 \over 3} \, c \, \sigma_T \, {L_{\rm D} \over 4 \pi 
\, m_e c^3 \, \Gamma_j^2 (1+ \beta_j)^2 \, x^{2}_0} =
\, {\sigma_T  L_{\rm D} \over 3 \pi \, m_e c^2 \, \Gamma_j^2 \, (1+\beta_j)^2 \, x^{2}_0}
\label{nu_D},
\end{equation}
and $f_{\ast} (t)$ is a function describing the decline of the stellar radiation 
field energy density with distance from the black hole. For the purpose
of an analytical description, we will approximate $f_{\ast} (t)$ as a broken
power-law 
\begin{equation}
f_{\ast} (t) = {\xi \over x^2 + s^2} \cong\cases{\xi \, s^{-2} & for $x^{} \le s$, \cr
\xi \, x^{-2} & for $x^{} \ge s$ \cr}
\label{bpl_star}
\end{equation}

where $\xi \cong \, \sigma_{\rm T} L_{\ast} \, \Gamma_j^2/3 \pi \,
m_e c^2$. An analytic solution to Eq. \ref{cooling2} will be derived
in the Appendix and discussed in \S \ref{analytic}.

Once a solution $\gamma_i (t_i; \gamma, t)$ to Eq. \ref{cooling2} is 
found, the electron distribution at any point in 
time (and thus at any point along the jet) can be calculated through
the expression

\begin{equation}
N_e (\gamma; t) = \int\limits_{t_0}^t dt_i \; {d^2 N_e (\gamma_i, t_i)
\over d\gamma_i \, dt_i} \, \left\vert{d\gamma_i \over d\gamma}
\right\vert.
\label{N_general}
\end{equation}
Additional restrictions on the integration in Eq. \ref{N_general}
and possible solutions of $\gamma (t; \gamma_i, t_i)$ are imposed
by the Heaviside functions in Eq. \ref{injection} and will be specified
in \S \ref{analytic}.

Radiation mechanisms included in our simulations are synchrotron emission, 
Compton upscattering of synchrotron photons, namely synchrotron self-Compton
(SSC) emission, and Compton upscattering of external photons. With the 
time-dependent (and thus $x$-dependent) non-thermal electron spectra in 
Eq. \ref{N_general}, we then use a $\delta$-function approximation to 
estimate the $\nu F_{\nu}$ synchrotron spectral output $\fesy$ at a 
dimensionless photon energy $\epsilon = h \nu / m_e c^2$ (in the observer's frame):
\begin{equation}
F_{\nu}^{syn}=\frac{1}{4 \pi \,d_L^2} \frac{h}{m_e c^2} \frac{d^2 E}{dt \, d{\epsilon}}
\end {equation}
 and 
\begin{equation}
\frac{d^2 E}{dt \, d{\epsilon}} = \frac{4}{3} \, c \sigma_T u_B \gamma_{sy}^2 N(\gamma)
\end {equation}

giving
\begin{equation}
\fesy (\epsilon, t_{obs}) = D^4 \left( {c \, \sigma_T \, u_B \over 
6 \pi \, d^2} \right) \gamma_{\rm sy}^3 \, N_e (\gamma_{\rm sy}, t)\;, \; \gamma_{\rm sy} 
\equiv \sqrt{ {\epsilon \over D \epsilon_B}}
\label{fsy}
\end{equation}
\citep{dss97}, where $D = 
\left[ \Gamma_j \, (1 - \beta_j \mu) \right]^{-1}$ is the Doppler 
boosting factor, $\epsilon_B \equiv B / B_{\rm cr}$ with 
$B_{\rm cr} = m^2 c^3/\hbar e = 4.414 \times 10^{13}$~G, defined as the field at which the 
cyclotron quantum number equals the rest mass energy of the electron, $t_{obs}$ is the
observer time, so that $ t_{obs} = t/D  = t^{\ast}_{\rm a, r} /D \Gamma_j$, and $E$ is the 
total synchrotron energy radiated in the system.

For the time-dependent $\nu F_{\nu}$ spectral output $\feec$ due to 
Compton upscattering of external photons from the star, we use the
Thomson approximation and represent the star as a monochromatic 
point source. We find
\begin{equation}
\feec(\epsilon, t_{\rm obs}) = D^4 \left( {c \, \sigma_T \, u_{ext} \over 
6 \pi \, d^2} \right) \gamma_{\rm c}^3 \, N_e (\gamma_{\rm c}, t)\;,\; 
\gamma_c \equiv \sqrt{{\epsilon\over D \epsilon_{ext}}},
\label{fec}
\end{equation}
where $u_{ext}$ is the Doppler-boosted seed photon energy density from 
the companion star (in the co-moving frame), given by
\begin{equation}
u_{ext} = D_{*}^2 \frac{L_{\ast}}{4\pi y^2 c},
\end{equation}
where $\epsilon_{\rm ext}$ 
is the dimensionless photon energy due to the companion star in the co-moving 
frame. It is related to the characteristic photon energy from the star $\epsilon_{\ast}
 = 2.7 \, kT_{\ast} / m_e c^2$ through $\epsilon_{ext} = D_{\ast} \epsilon_{\ast}$. $T_{\ast}$ 
is the effective surface temperature of the star, and $D_{\ast} = 
\Gamma_j(1 - \beta_{\Gamma} \mu_{ext}^{\ast})$ Lorentz transforms $\epsilon_{\ast}$ 
into the comoving frame.

Here we have omitted an angle-dependent factor that arises in a 
more accurate derivation for the spectrum of Thomson-scattered stellar radiation, which
is important when the orbital variability of the star is considered \citep{db06}.

The photon spectrum due to Compton upscattering of photons from the accretion 
disk is obtained by adopting a similar formalism as that used for the companion star. 
We approximate the disk as a thin annulus at the radius $R_{\rm max}$ where 
the differential energy output is maximized. We use the optically thick, 
geometrically thin, gas-pressure dominated accretion disk solution of 
\cite{ss73} predicting a blackbody spectrum according to a temperature 
distribution $T_D(R_{max})$ given by
\begin{equation}
\Theta(R_{max}) =\frac{kT_D}{m_ec^2}(R_{max})
\end{equation}
\begin{equation}
= 1.44 \left(\frac{M}{M_\odot}\right)^{-\frac{1}{2}}
\left(\frac{\dot M}{M_\odot/yr}\right)^{\frac{1}{4}}
\left(\frac{R_{max}}{R_g}\right)^{-\frac{3}{4}}
\left(1-\sqrt{\frac{6R_g}{R_{max}}}\right)^{\frac{1}{4}}
\end{equation}
where $R_g$ is the gravitational radius, $R_{\rm max} = 9.375 \, R_g$, $T_D$ is 
the corresponding disk temperature, $M$ is the black hole mass in units of 
$M_\odot$, and $\dot M$ is the mass accretion rate.
 
We also include SSC emission, keeping in mind that our 
analytical approach outlined in the following section will not be applicable to 
situations in which the radiative output (and thus electron cooling) from SSC 
dominates the bolometric luminosity, since we are neglecting SSC cooling of the
electrons. We use the formalism of \cite{tavecchio98} to calculate the SSC 
spectral output in the Thomson regime. In order to be able to use this formalism,
we approximate our time-dependent population of relativistic electrons at any given
time by a broken power law distribution with indices $n_{1}$ and $n_{2}$ and break 
Lorentz factor $\gamma_b$:
\begin{equation}
N(\gamma) = \left\{ \begin{array}{ll}
      K_{1}\gamma^{-n_{1}} & \mbox{if $\gamma<\gamma_{b}$}
\\    K_{2}\gamma^{-n_{2}} & \mbox{if $\gamma_{b}<\gamma<\gamma_{max}$}, \end{array} \right.
\end{equation} 
where the coefficients $K_{1}$ and $K_{2}$ are related through the kinetic condition 
$K_{1}\gamma_{b}^{-n_{1}} = K_{2}\gamma_{b}^{-n_{2}}$.
The value of $\gamma_b$ is determined by the solution of the kinetic equation 
for electron cooling.
For weakly beamed jet emission, some fraction of the radiation is expected to impinge upon 
the cooler material in the accretion disk and lead to fluorescent Fe K$\alpha$ line emission 
and a reflection hump \citep{beloborodov99}. The amount of radiation irradiating the disk 
depends on the distance between the jet emission zone and the disk, the jet inclination angle, 
and the bulk Lorentz factor $\Gamma_j$. X-ray emission originating at distances $\sim 10^3 R_g$ 
from the disk, together with $\beta_j=0.92$ in our model, corresponds to the case of a 
``synchrotron dominated jet'' \citep{mn04,mnw05}, where the fraction 
of reflected emission has been calculated to be only $\approx$ 1$\%$-2$\%$. We therefore 
do not include this component in our present calculations.

\section{\label{analytic}Analytic Solutions to the Electron Kinetic Equation}

The dynamic equation (\ref{cooling2}) can be solved analytically for arbitrary 
values of $a$, as long as the distance dependence of the external radiation
fields, characterized by $f_{\ast} (t)$, can be piecewise approximated as a 
power-law. The solution to the electron cooling problem, as derived in the Appendix, is given by
$$\gamma (t; \gamma_i, t_i) = t^{-m} \Big[ {1 \over \gamma_i t_i^m} 
+ {\beta \over 1 - m - 4a} \left( t^{1 - m - 4a} - t_i^{1 - m - 4a} \right) + $$
\begin{equation}
 {\delta \over 1 + m} \left( t_i^{-(1 + m)} - t^{-(1 + m)} \right) 
+ \int_{t_i}^{t} f_{\ast} (t') \, {t'}^{-m} \, dt' \Big]^{-1} 
\label{general_g}
\end{equation}
where
$\beta = \nu_0 t_0^{4a}$ and $\delta = \nu_{\rm D} t_0^2$. Note that one can
easily incorporate as many additional external photon components with different
decay slopes and normalizations as required for any specific problem at hand. 

The evaluation of the time-dependent electron distribution becomes more
easily tractable if we transform the $t_i$ integration in Eq.\ \ref{N_general} 
into an integral over the injection energy $\gamma_i$:
\begin{equation}
N_e (\gamma, t) = Q_0 \int\limits_{\gimin}^{\gimax} d\gamma_i \;
\gamma_i^{-q} \, \left\vert {d\gamma_i \over d\gamma} \right\vert
\, \left\vert {dt_i \over d\gamma_i} \right\vert
\label{N_gen_app}
\end{equation}
The expression for the Jacobian in Eq. \ref{N_gen_app} based on our 
analytical solution is given in the Appendix (Eq. \ref{jacobian_star}). 
The boundaries of the integral (\ref{N_gen_app}) follow from the
Heaviside functions in Eq. \ref{N_general}, which also provide
the relevant limits on the values of $\gamma$ for which the electron
distribution is non-zero. Depending on whether the emission region 
is currently within the zone of electron injection ($t \le t_1$) or 
beyond it ($t \ge t_1$), we find two sets of boundary conditions.

In the case $t \le t_1$, we have 
\begin{equation}
\begin{array}{rl}
\gimin &= \gamma, \cr
\gimax &= \min \left( \gmax, \gamma_i [t_0; \gamma, t] \right), \cr
\end{array}
\label{gilimit1}
\end{equation}
with non-zero electron distribution for
\begin{equation}
\gamma (t; \gmin, t_0) \le \gamma \le \gmax
\label{glimit1}
\end{equation}
where $\gamma_i [t_0; \gamma, t]$ is found by inversion of solution
in Eq. \ref{general_g}. 

In the case $t \ge t_1$, we find
\begin{equation}
\begin{array}{rl}
\gimin &= \max \left( \gmin, \gamma_i [t_1; \gamma, t] \right), \cr
\gimax &= \min \left( \gmax, \gamma_i [t_0; \gamma, t] \right), \cr
\end{array}
\label{gilimit2}
\end{equation}
with non-zero electron distribution for
\begin{equation}
\gamma (t; \gmin, t_0) \le \gamma \le \gamma(t; \gmax, t_1).
\label{glimit2}
\end{equation}

The solution, Eq.\  \ref{general_g}, reduces to a particularly simple 
form in the limit of negligible radiative losses, in which the electron 
kinetic equation reduces to $\dot\gamma = - m (\gamma / t)$, and 
$\gamma(t; \gamma_i, t_i) = \gamma_i \, (t / t_i)^{-m}$. The local 
electron spectrum is then given by 
\begin{equation}
N_e^{\rm adi} (\gamma; t) = Q_0 \, \gamma^{-q} \, t \, {\left( 
{t_{\rm i, max} \over t} \right)^{\eta} - \left( {t_{\rm i, min} 
\over t} \right)^{\eta} \over \eta}
\label{Ne_adi}
\end{equation}
where $t_{\rm i, min}$ and $t_{\rm i, max}$ are determined by the 
Heaviside function in Eq. \ref{injection}, and $\eta = (q - 1) \, 
m + 1$. In the limit $t \gg t_1$ and $\gamma \ll \gamma_{\rm max} (t)$, 
the local electron spectrum reduces to
\begin{equation}
N_e^{\rm adi} (\gamma; t) \approx {Q_0 \, t_1 \over \eta} \,
\gamma^{-q} \, \left( {t \over t_1} \right)^{1 - \eta}.
\label{Ne_adi_approx}
\end{equation}

\begin{deluxetable}{cccccccccc}
\tabletypesize{\scriptsize}
\tablecaption{Parameter choices for our parameter study}
\tablewidth{0pt}
\tablehead{
\colhead{Sequence} & \colhead{$B_0$ [G]} & \colhead{$L_{\ast}$ [ergs~s$^{-1}$]} & 
\colhead{$q$} & \colhead{$\gamma_1$} & \colhead{$\gamma_2$} & 
\colhead{$L_{\rm inj}$ [L$_{Edd}$]} & \colhead{$\theta_{\rm obs}$ [$^o$]} & 
\colhead{$D$}
}
\startdata
(1) & $1.0 \times 10^3$ & $8 \times 10^{37}$ & 2.4 & 10 & $10^4$ & $4.4 \times 10^{-5}$ & 70 & 0.583 \\
(1) & $2.5 \times 10^3$ & \\
(1) & $5.0 \times 10^3$ & \\
(1) & $7.5 \times 10^3$ & \\
(1) & $1.0 \times 10^4$ & \\
\noalign{\smallskip\hrule\smallskip}
(2) & $5.0 \times 10^3$ & $8 \times 10^{36}$ & 2.4 & 10 & $10^4$ & $4.4 \times 10^{-5}$ & 70 & 0.583 \\
(2) &                   & $8 \times 10^{37}$ & \\
(2) &                   & $8 \times 10^{38}$ & \\
(2) &                   & $8 \times 10^{39}$ & \\
\noalign{\smallskip\hrule\smallskip}
(3) & $5.0 \times 10^3$ & $8 \times 10^{36}$ & 1.2 & 10 & $10^4$ & $4.4 \times 10^{-5}$ & 70 & 0.583 \\
(3) &                   &                    & 1.8 & \\
(3) &                   &                    & 2.4 & \\
(3) &                   &                    & 2.8 & \\
\noalign{\smallskip\hrule\smallskip}
(4) & $5.0 \times 10^3$ & $8 \times 10^{36}$ & 2.4 & 1  & $10^4$ & $4.4 \times 10^{-5}$ & 70 & 0.583 \\
(4) &                   &                    &     & 10 & \\
(4) &                   &                    &     & 100 & \\
(4) &                   &                    &     & 1000 & \\
\noalign{\smallskip\hrule\smallskip}
(5) & $5.0 \times 10^3$ & $8 \times 10^{36}$ & 2.4 & 10 & $10^4$ & $4.4 \times 10^{-5}$ & 70 & 0.583 \\
(5) &                   &                    &     &    & $10^5$ \\
(5) &                   &                    &     &    & $10^6$ \\
\noalign{\smallskip\hrule\smallskip}
(6) & $5.0 \times 10^3$ & $8 \times 10^{36}$ & 2.4 & 10 & $10^4$ & $2.2 \times 10^{-5}$ & 70 & 0.583 \\
(6) &                   &                    &     &    &        & $4.4 \times 10^{-5}$ \\
(6) &                   &                    &     &    &        & $2.2 \times 10^{-4}$ \\
(6) &                   &                    &     &    &        & $4.4 \times 10^{-4}$ \\
\noalign{\smallskip\hrule\smallskip}
(7) & $5.0 \times 10^3$ & $8 \times 10^{36}$ & 2.4 & 10 & $10^4$ & $1 \times 10^{35}$ & 5  & 4.60 \\
(7) &                   &                    &     &    &        &                    & 20 & 2.88 \\
(7) &                   &                    &     &    &        &                    & 35 & 1.60 \\
(7) &                   &                    &     &    &        &                    & 70 & 0.583 \\
\enddata
\label{parstudy_table}
\end{deluxetable}

\section{\label{parameterstudy}Results}

A large number of simulations have been performed to study the effects of the various model
parameters on the resulting broadband spectra, light curves, and X-ray hardness intensity 
diagrams (HIDs). We start our parameter study with a baseline model for which we have 
used the standard model parameters discussed in \S~\ref{model} and listed in Table~\ref{par_table}. 
Subsequently, we investigate the departure from this standard set-up by varying (1) the initial 
magnetic field $B_0$, (2) the luminosity of the companion star $L_{\ast}$, (3) the injection 
electron spectral index $q$, (4) the low-energy cutoff $\gamma_1$ of the electron injection 
spectrum, (5) the high-energy cutoff $\gamma_2$ of the electron injection spectrum, (6) the 
injection luminosity $L_{inj}$, (7) and the observing angle $\theta_{obs}$ and thus the Doppler 
boosting factor. The parameters used for the individual runs are quoted in Table \ref{parstudy_table}. 

\begin{figure}[t]
\includegraphics[height=12cm]{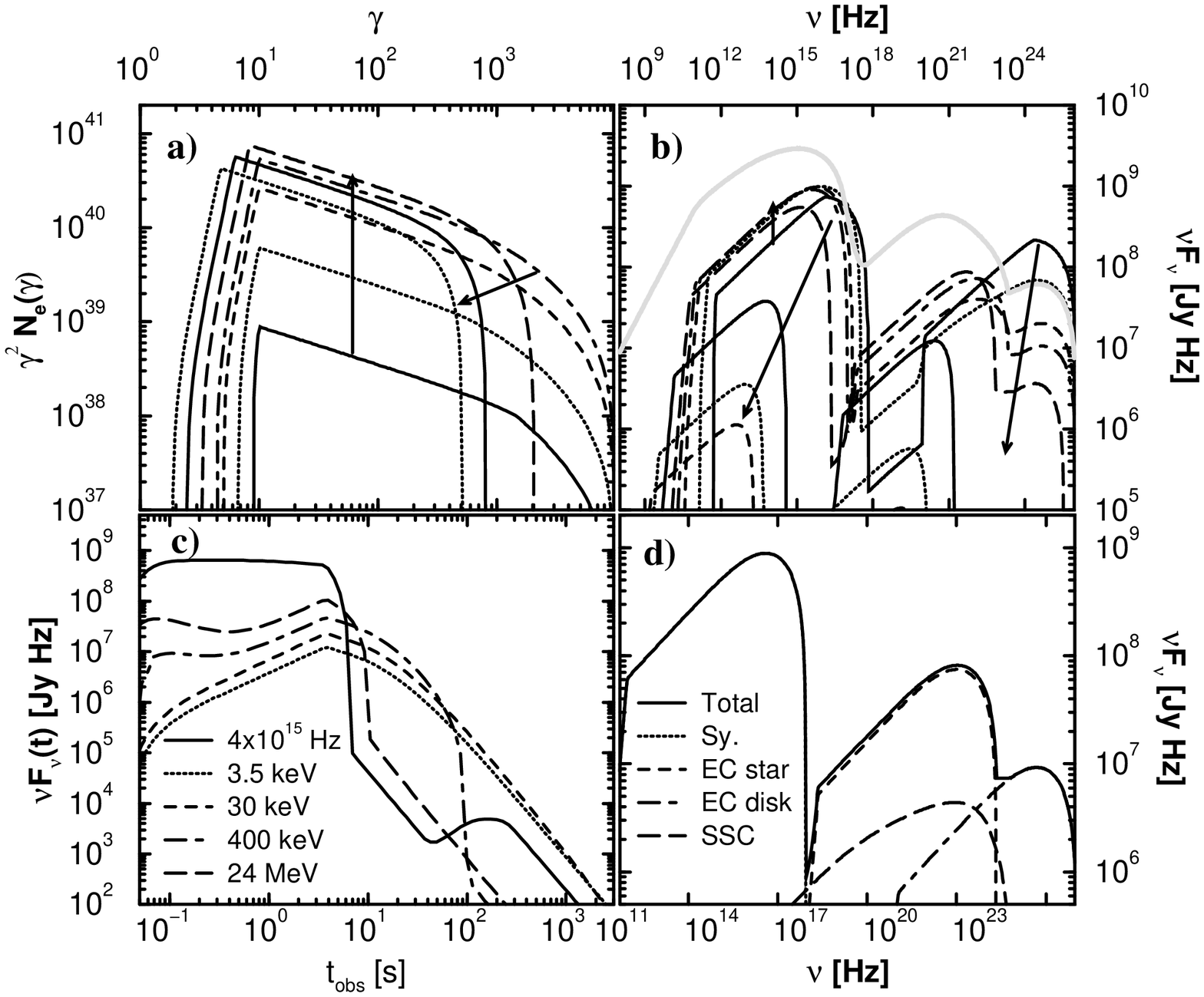}
\caption{The baseline model (for parameters see Table \ref{par_table}): (a) Time-dependent
electron spectra, beginning at $x_0$, increasing by a multiplicative step of 10
between individual curves. The arrows indicate the sense of time evolution; the last step
corresponds to a co-moving time of 87.6~s. (b) Time-dependent $\nu F_{\nu}$ photon spectra,
corresponding to the same times as shown in panel a. The heavy gray curve is the time-integrated
flux (fluence, in units of Jy~Hz~s). (c) Light curves at five frequencies, as indicated in
the legend. (d) Snap-shot spectrum just before the end of the injection period, corresponding
to $t \sim 1.5$~s, showing the individual emission components.}
\label{baseline}
\end{figure}

\begin{figure}[t]
\includegraphics[height=12cm]{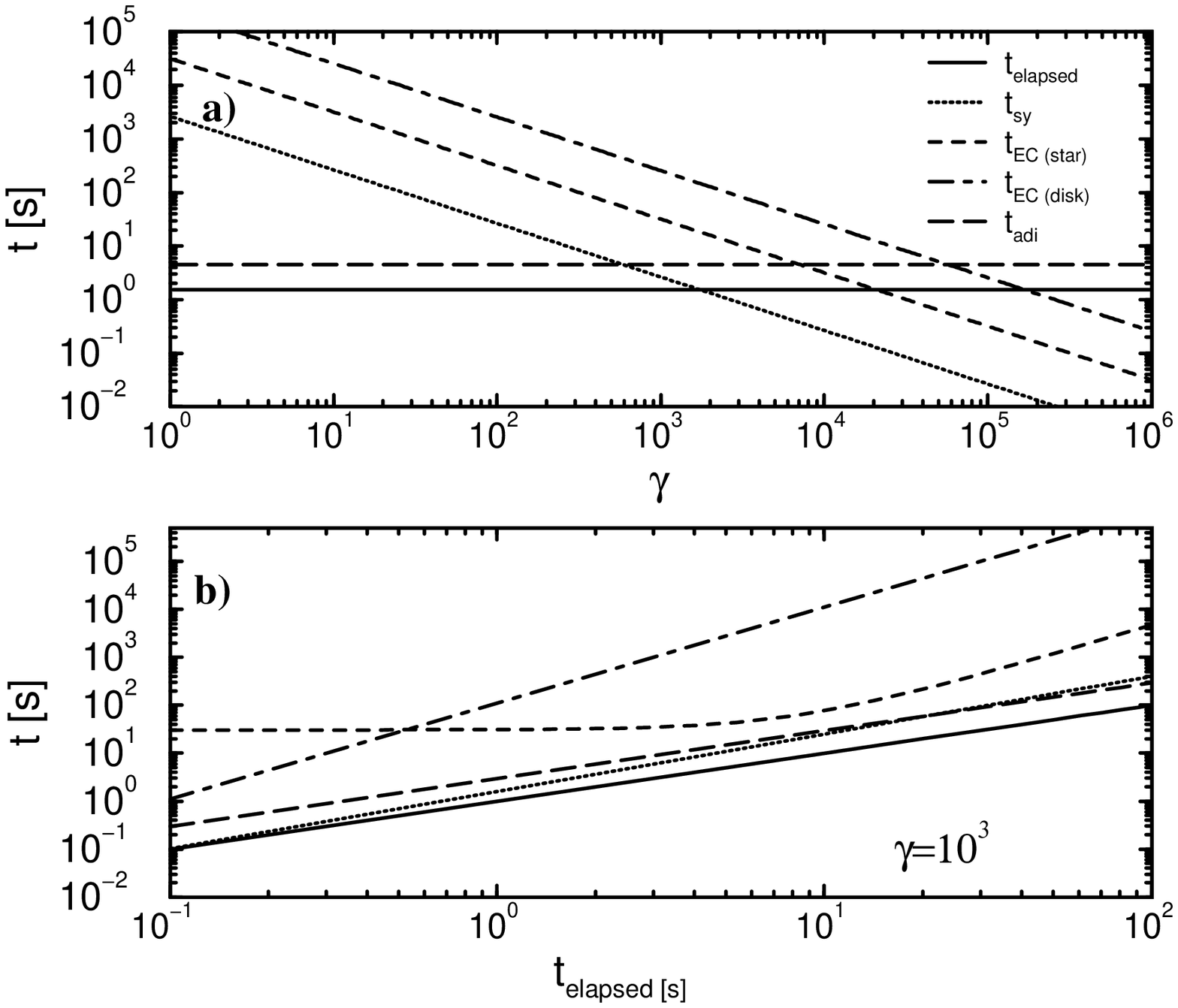}
\caption{Electron cooling time scales for our baseline model (a) as a function of 
electron energy right before the end of the injection period (same time as in panel 
\ref{baseline}d), and (b) as a function of time for a fixed electron energy of 
$\gamma = 10^3$.}
\label{cooling}
\end{figure}

In Figure \ref{baseline}, we have compiled for our baseline model (a) a sequence of comoving 
electron spectra, (b) snapshot SEDs, and the time averaged photon spectrum, (c) light curves 
at various photon energies, and (d) a snap-shot spectrum, indicating the individual emission
components. Fig. \ref{baseline}a illustrates the gradual build-up of the electron density in 
the emission region and their subsequent radiative and adiabatic cooling. Fig. \ref{cooling}
illustrates the electron cooling time scales (a) as a function of electron energy at the end
of the injection period, and (b) as a function of time for a fixed electron energy of
$\gamma = 10^3$. At the end of the injection period, the radiative cooling timescale is 
shorter than the elapsed time for electron energies of $\gamma \ge \gamma_b = 2.4 \times 
10^3$. In our baseline model, the magnetic field energy density is higher than that of the 
external photon field, with
\begin{equation}
\frac{u_{ext}}{u_B}\sim\frac{2 \, D_*^2 \, L_{\ast}}{B^2 \, y^2 \, c}
\end{equation}
attaining a value of 0.36 at the end of the injection period. Because of the rapid decline
of the magnetic field, $B \propto x^{-2a} \propto x^{-0.6}$, the cooling time scale 
increases with time in our base model, as $t_{\rm sy} \propto t^{4a} \propto t^{1.2}$. 
For this reason, one can see a synchrotron cooling break in the electron spectra around
$\gamma_b \sim 10^3$ at early times, while, at the end of the injection period, this
break blends in with a gradual high-energy cut-off and is no longer discernible. After 
the end of the injection period, synchrotron cooling is initially still dominant, but 
adiabatic cooling is gradually taking over as the dominant energy loss mechanism (see
Fig. \ref{cooling}b). 

The time-dependent photon spectra in Fig. \ref{baseline}b indicate that the synchrotron 
and the Compton (disk) components decay even within the injection interval due to decaying 
source fields. The disk photon energy density begins to decline ($\propto t^{-2}$) 
immediately after the onset of injection, while the stellar photon field remains basically 
constant for $x < s$. This manifests itself in the gradual dominance of the 
EC (star) component over the EC (disk) component in its time evolution. The EC (disk) 
component produces the highest-energy photons, extending out to energies beyond 
$\sim 100$~GeV, but only during a very short, sub-second flash very close to $x_0$. 
If the electron injection does actually start as close to the black hole as $ x_0 
\sim 1000 \rm R_g$ , as we have assumed here, such high-energy radiation might suffer 
substantial $\gamma\gamma$ absorption in the intense radiation field of the stellar companion
\citep{moskalenko93,moskalenko94,bednarek00,boettcher05} and might thus not be directly 
observable. However, as noted in the introduction, Compton scattering of the stellar radiation
field or possibly also hadronic processes might very well produce VHE $\gamma$-rays at 
larger distances from the central compact object \citep{romero03,bosch05a,db06} 
which could escape the compact radiation field of the stellar companion without 
significant $\gamma\gamma$ attenuation. 
The radio spectrum is usually flat to inverted and extends to the infrared and 
even beyond \citep{fender00}. The spectral cut-off above the GHz range in  
Fig.~\ref{baseline}b is a result of our choice of $\gamma_1=10$, the lower
energy cut-off of the electron distribution. Effects of varying  $\gamma_1$ are
investigated in Section 4.5, showing a spectral shift towards lower frequencies 
with lower values of $\gamma_1$. The short-time X-ray spectral variability predictions, 
which are the main focus of this paper, are only weakly dependent on the choice 
of $\gamma_1$, at least within a realistic range of $\gamma_1\sim 1-10$.

We also note (see Fig.~\ref{baseline}d) that the contribution from the SSC 
photons to the total photon spectra is negligible in our base model. 

Figure \ref{baseline}c shows light curves at different photon frequencies, one in the UV,
two in X-rays, and two at $\gamma$-ray energies. From the light curves and the snap-shot
spectra in Fig. \ref{baseline}b, one can see that the synchrotron-dominated optical -- UV 
spectra reach their maximum very rapidly and remain at an approximately constant level 
until the higher-energy light curves reach their maxima, around the end of the injection
period. The approximately constant level of the optical -- UV radiation results from the 
opposing effects of an increasing electron density and a decreasing magnetic field. 
There is no significant time delay between the light curve peaks at the various X-ray
and $\gamma$-ray energies. At 24 MeV, the dip in the light curve before the end of injection 
is a result of the very rapid decay of the photon field density from the disk. Here, the 
EC (disk) component dominates very early on, but declines rapidly and becomes dominated 
by the rising EC (star) component within less than 1~s. In general, a rapid change in 
temporal slope, as seen in various light curves in Fig. \ref{baseline}c, indicates a 
transition from one radiation component to another, passing through the observing range. 
It should be mentioned here that the hard low-frequency cut-off apparent in our 
photon spectra instead of a $\nu F_{\nu}^{syn}\propto{\nu^{4/3}}$ behavior at 
$\nu<\nu (\gamma_1)$ is a result of the $\delta$-function approximation in our 
calculations.

In the following parameter study, we focus on the time-averaged photon spectra, light 
curves, and X-ray HIDs, and explore the effect of variations of individual parameters 
on these aspects. Throughout our study, the dominant radiative cooling mechanism will 
either be synchrotron or EC (star). We find that for plausible choices of parameters 
the contribution from SSC cooling will always be negligible, as required by our 
analytical approach to the solution of the electron kinetic equation given in 
Eq. (\ref{general_g}).

\subsection{\emph {Initial Magnetic field}}

\begin{figure}[t]
\includegraphics[height=12cm]{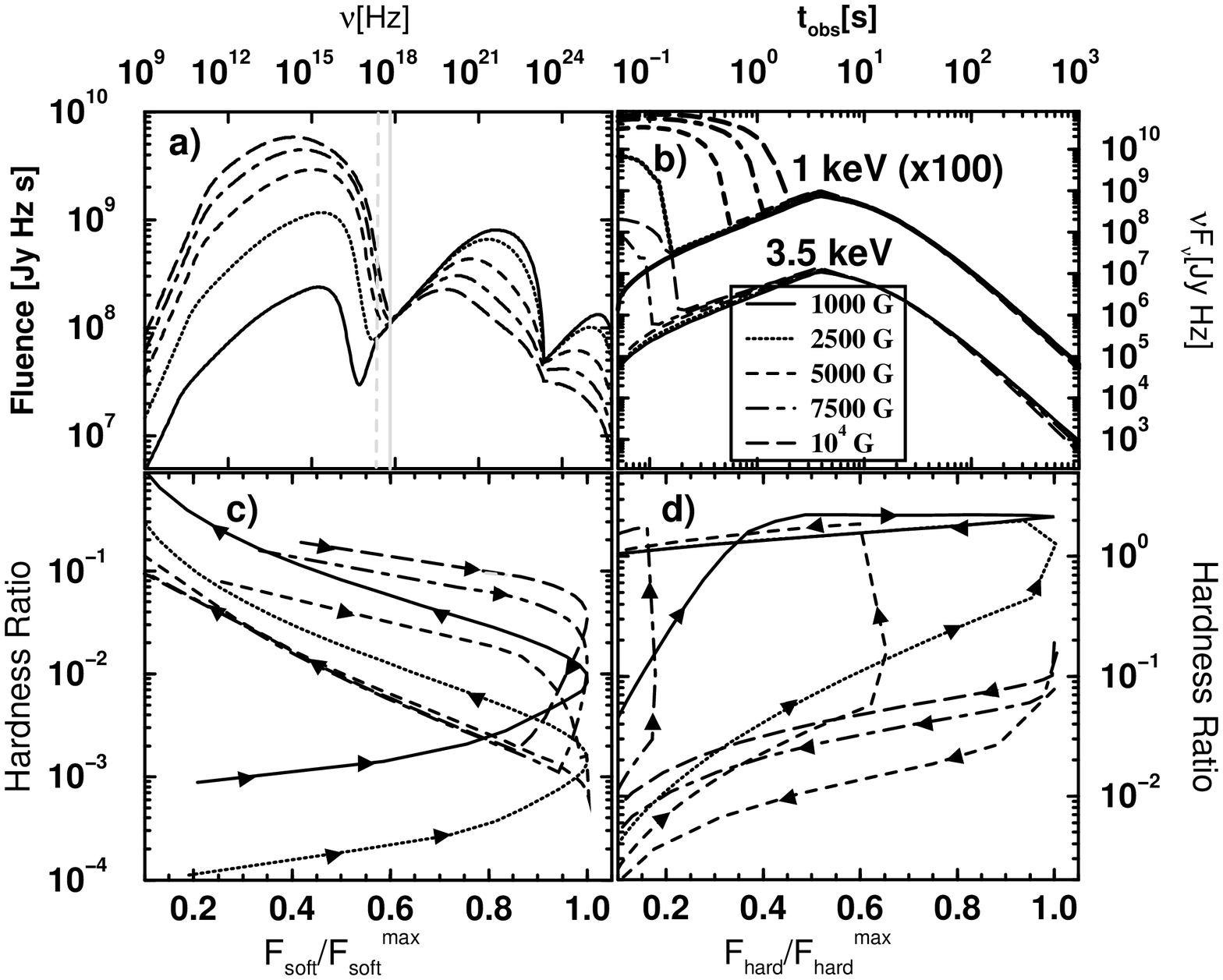}
\caption{Effect of a changing magnetic field on (a) the time-integrated $\nu F_{\nu}$ (fluence)
spectra, (b) the X-ray light curves at 1~keV (multiplied by 100 for clarity) and 3.5~keV, and 
(c,d) hardness-intensity diagrams (HIDs). In the HIDs, the soft flux is the integrated 0.1 -- 
2~keV flux, the hard flux is the 2 -- 10~keV flux, and the hardness ratio is the ratio of the 
two. The abrupt shape of some of the HID tracks is an artifact of the $\delta$ approximations
used for some of our spectral calculations. The respective magnetic fields are indicated in 
the legend; the other parameters are the baseline model values discussed in the text and 
listed in Table \ref{par_table}. The vertical lines in panel (a) indicate the photon energies
at which the light curves in panel (b) were extracted.}
\label{Bchange}
\end{figure}

In this subsection, we investigate how different choices of the initial magnetic field
$B_0$ (at $x_0$) influence the shape of the time-averaged photon spectra, light curves, 
and X-ray HIDs. The results are illustrated in Fig. \ref{Bchange}.

With increasing values of $B_0$, obviously, the total output in synchrotron emission increases,
at the expense of the Compton components. As all other parameters remain constant, this also
leads to a more rapid radiative cooling of the electrons, resulting in a cooling break at lower
electron energies and, consequently, lower $\nu F_{\nu}$ peak frequencies of all radiation 
components. Notice that the time-averaged flux at $\sim 10$~keV remains almost unchanged for 
different magnetic field values. For the parameters adopted here, the X-ray emission is generally
dominated by the EC (star) component, except for very high magnetic field values, in which case
the synchrotron component initially extends into the X-ray regime. This is reflected in the
very high X-ray fluxes in the highest-magnetic-field cases (see Fig. \ref{Bchange}b). Note that
in the Compton-dominated X-ray regime (3.5~keV), the effect of different magnetic-field values 
on the light curve is only very minor. This is because this part of the spectrum is dominated
by the low-energy end of the EC (star) component, produced by low-energy electrons, whose cooling
is dominated by adiabatic cooling (see Fig. \ref{cooling}a), irrespective of the magnetic field.

The shorter radiative cooling time scale with increasing $B_0$ (and thus decreasing $<\gamma^2>$)
also leads to an even further decreasing contribution of the SSC component since 
\begin{equation}
\frac{L_{ssc}}{L_{syn}}\sim<\gamma^2>\sigma_T \, n_e \, R.
\label{Lssc}
\end{equation} 

Panels c) and d) compile the hardness intensity diagrams for the different values of $B_0$. 
Here and throughout the rest of the paper, we define the hardness ratio as the ratio of the 
X-ray fluxes at 2 -- 10~keV (hard X-ray flux) to 0.1 -- 2~keV (soft X-ray flux). The abrupt 
breaks in some of these tracks are an artifact of the $\delta$ approximations used in some 
of our spectral calculations. The HIDs trace out 
characteristic hysteresis loops, changing their orientation. This behaviour is expected if the
main contribution of the flux in a given energy bin transits between the high-energy end of
one emission component to the low-energy end of another one \citep[in this case, between the 
synchrotron and EC (star) components; see][]{bc02}. At X-ray energies around the synchrotron 
cut-off, the flux maxima occur at significantly different hardness ratios for different 
values of $B_0$ (Fig. \ref{Bchange}d). Specifically, within the synchrotron-dominated,
clockwise spectral hysteresis loops, the peak flux is accompanied by a harder spectrum
for higher magnetic fields. This is a consequence of the increasing frequency of the
synchrotron peak with increasing magnetic field. Within the Compton-dominated, counterclockwise 
spectral hysterersis loops, the peak hard X-ray flux is accompanied by a softer hardness ratio
for larger values of $B_0$, which is a consequence of the more efficient electron cooling. 

\subsection{\emph {Luminosity of the Companion Star}}

\begin{figure}[t]
\includegraphics[height=12cm]{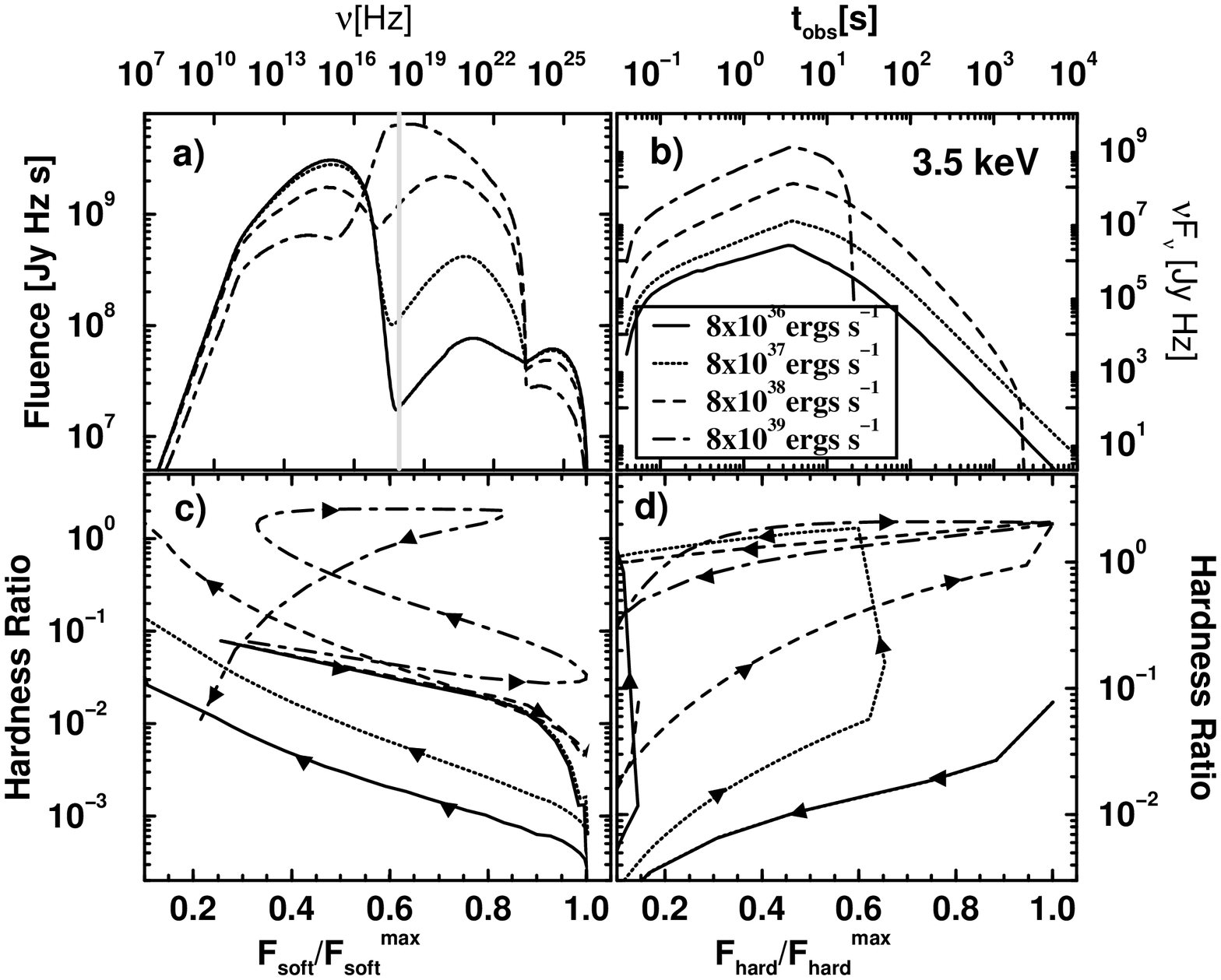}
\caption{Effect of a changing luminosity of the companion star on (a) the time-integrated 
$\nu F_{\nu}$ (fluence) spectra, (b) the X-ray light curves at 3.5~keV, and 
(c,d) hardness-intensity diagrams (HIDs). The respective values of $L_{\ast}$ are indicated 
in the legend; the other parameters are the baseline model values discussed in the text and 
listed in Table \ref{par_table}. The vertical line in panel (a) indicates the photon energy
of 3.5~keV for which the light curves in panel (b) are extracted. }
\label{Lchange}
\end{figure}

In our base model, as in most cases with substantial mass transfer rates onto the compact
object, the external source photon field for Compton scattering is dominated by the star
light of the stellar companion. For this reason, we performed a series of simulations with 
increasing values of the stellar luminosity to investigate the influence of a varying 
external photon density on the broadband spectra, light curves, and spectral variability
patterns. The results of these simulations are illustrated in Fig. \ref{Lchange}.

The time averaged photon spectra are shown in Fig. \ref{Lchange}a, which clearly shows the 
emergence of the EC (star) component at X-ray and $\gamma$-ray energies. The impact of an 
increasing star luminosity on other spectral components (at radio, optical, UV, and GeV 
$\gamma$-rays) is small as long as the synchrotron cooling time-scale is shorter than the 
EC (star) cooling time-scale (solid and dotted curves). When EC cooling becomes dominant 
over synchrotron cooling, it leads to a reduced power output in the synchrotron and EC (disk)
components as well as a shift of the $\nu F_{\nu}$ peak frequency of all radiation components
towards lower values (dashed and dot-dashed curves).

Fig. \ref{Lchange}b shows that the impact of a strong EC (star) component is quite prominent 
in the light curve at keV energies. Along with an increase in the peak $\nu F_{\nu}$ flux 
(Compton peak), an increase in the star luminosity also leads to a faster decay of the light 
curves in the high-luminosity cases, where the star light radiation energy density dominates 
over the magnetic-field energy density. 

The HIDs are plotted in Fig. \ref{Lchange}c,d and indicate X-ray spectral hysteresis. 
For low values of $L_{\ast}$, the X-ray fluxes are initially dominated by the synchrotron 
flux, leading to very soft spectra at the time of the peak soft flux. As the flux decays 
after the end of the injection period, the EC (star) component begins to sweep through 
the X-ray bands, producing increasingly hard spectra during the decaying portions of 
the light curves (i.e., the low-flux portions of the HIDs). As $L_{\ast}$ increases 
towards very high values, a secondary, counterclockwise loop in the soft-flux HID 
(Fig. \ref{Lchange}c) emerges when the EC (star) component sweeps through the X-ray 
bands. In that case, the EC (star) component dominates the hard X-ray flux very early 
on, causing the overall pattern of the hard flux HIDs (Fig. \ref{Lchange}d) to consist 
only of this counterclockwise loop. We conclude, that the observation of such a 
counterclockwise spectral hysteresis is a diagnostic of the dominance of the EC 
component in the X-ray regime. 

\subsection{\emph {Electron Injection Luminosity}}

\begin{figure}[t]
\includegraphics[height=12cm]{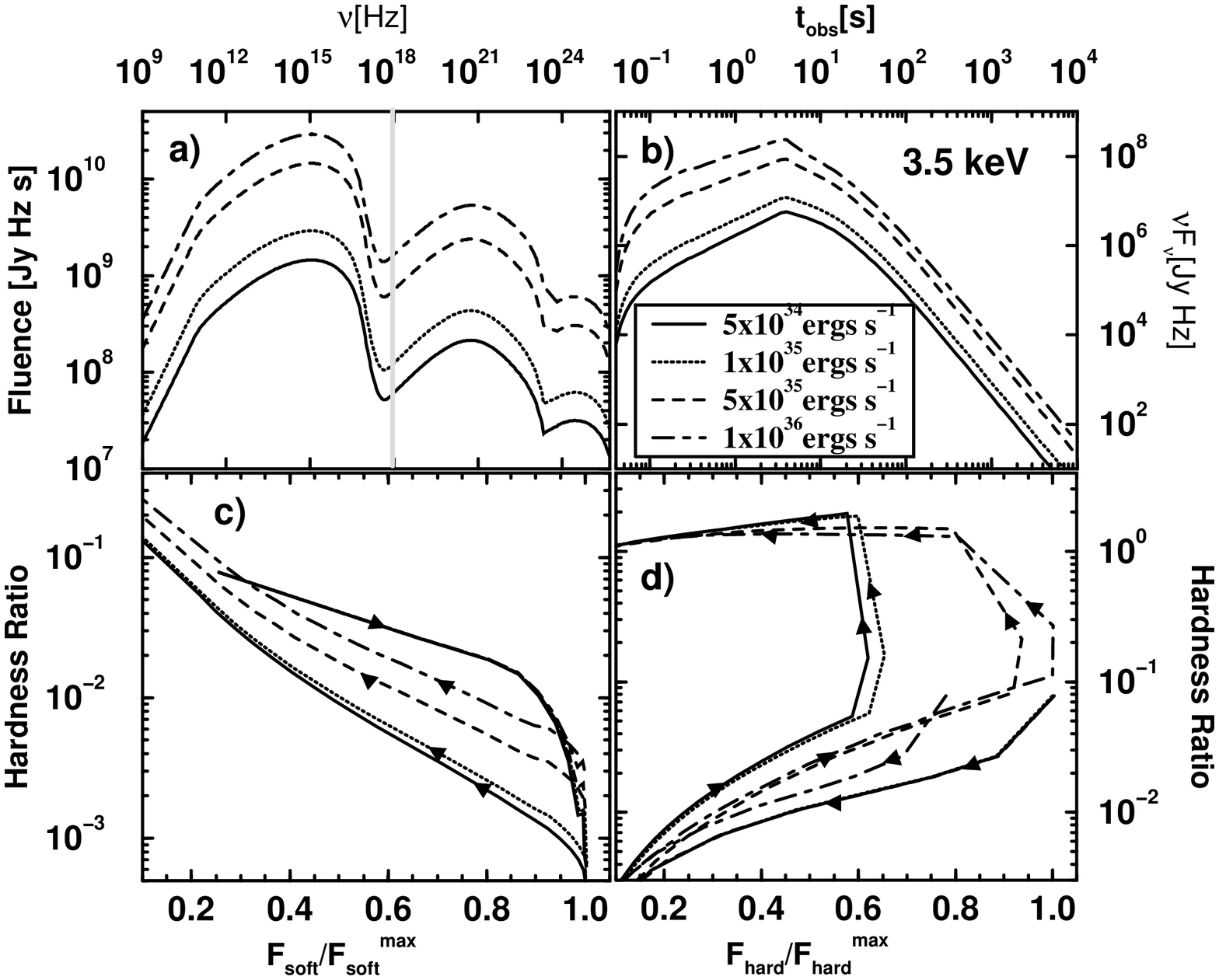}
\caption{Effect of a changing electron injection luminosity (see legend) on (a) the 
time-integrated $\nu F_{\nu}$ (fluence) spectra, (b) the X-ray light curves at 3.5~keV, 
and (c,d) hardness-intensity diagrams (HIDs). See Figs. \ref{Bchange} and \ref{Lchange}
for more explanations.}
\label{Lichange}
\end{figure}

The effect of an increasing injection luminosity, corresponding to a higher density of injected 
relativistic particles in the emitting region, is illustrated in Fig. \ref{Lichange}. This leads 
to a corresponding increase in the overall bolometric luminosity and a stronger energy output in 
the SSC component, relative to the synchrotron, as given in Equation \ref{Lssc}.

As long as the SSC emission is not dominant, this essentially only leads to a higher flux level
in the spectra and light curves. At high values of $L_{\rm inj}$, the SSC emission begins to
play a non-negligible role in the X-ray regime, which leads to a slightly shallower slope of the 
X-ray light curves during the rising phase ($t < t_1$). Also, for distances $x < s$, 
the decay slope of the X-ray light curve is steeper for higher values of $L_{\rm inj}$. Both of
these effects are a consequence of the fact that the magnetic field decay is faster than that of 
the star photon field, and as a consequence, the decay in seed photon density for the SSC emission 
is faster than that for the EC (star) emission. 

Panels c and d of Fig. \ref{Lichange} show the various HID tracks for different injection 
luminosities. The main effect at the soft band HIDs (Fig. \ref{Lichange}c) is visible in
a slight hardening of the spectra during the decay phase for high values of $L_{\rm inj}$
as the SSC emission gradually begins to play a non-negligible role. In the hard band
HIDs (Fig. \ref{Lichange}d), one can see that for low $L_{\rm inj}$, the luminosity peak
is reached very early on, and is dominated by the synchrotron component leaking into the
2 -- 10~keV band. A secondary counterclockwise loop emerges as the EC (disk) component
sweeps through the 2 -- 10~keV band. As $L_{\rm inj}$ increases (along with the SSC
contribution), the secondary loop extends towards higher relative flux values, until,
for the highest value of $L_{\rm inj}$, the peak flux occurs as the HID tracks goes
through the counterclockwise hysteresis loop. There is no significant difference in 
the local spectral index (or the hardness ratio) at the time of peak flux of the 
Compton-dominated hysteresis loop.

\subsection{\emph {Injection Electron Spectral Index}} 

\begin{figure}[t]
\includegraphics[height=12cm]{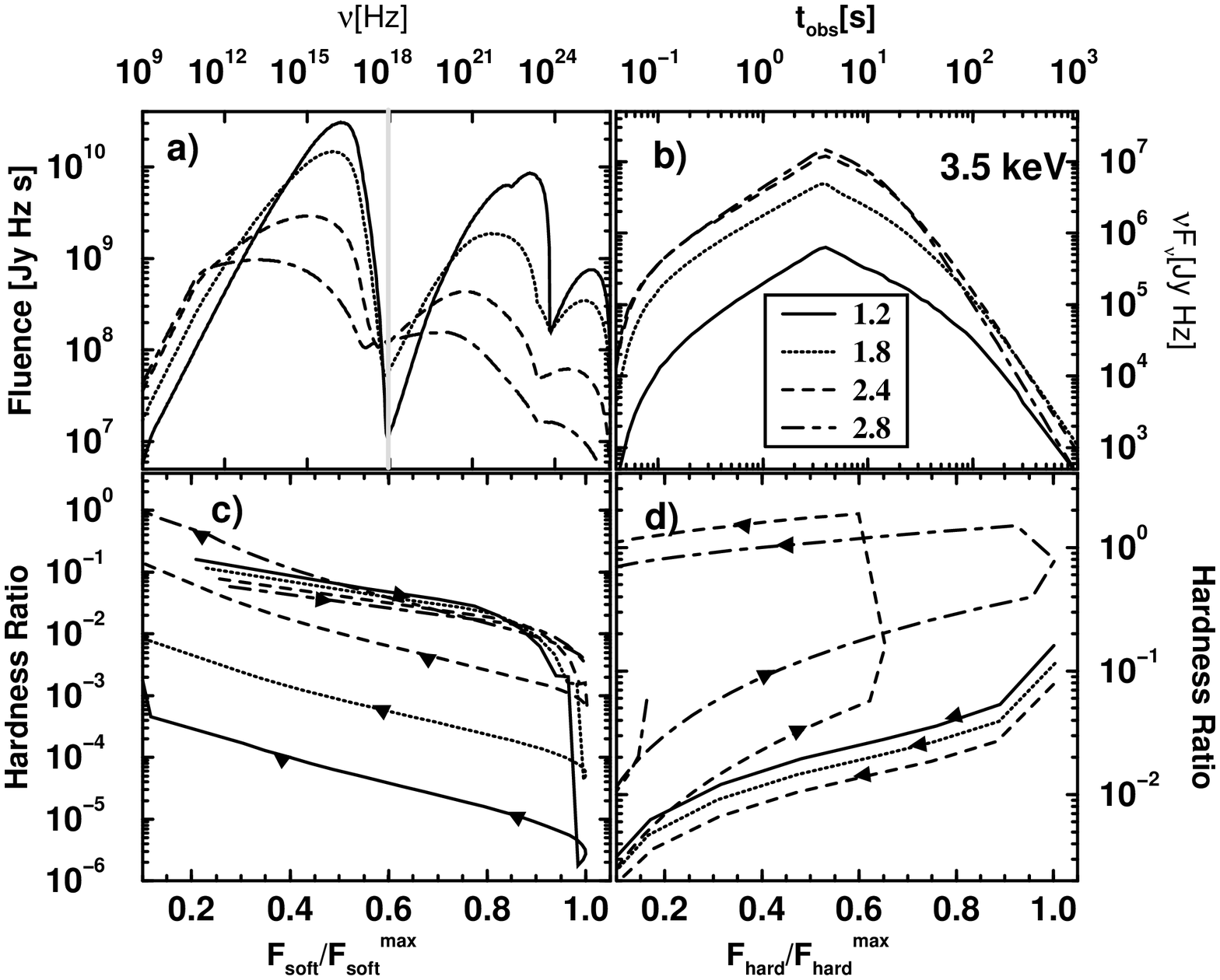}
\caption{Effect of a changing electron injection spectral index (see legend) on (a) the 
time-integrated $\nu F_{\nu}$ (fluence) spectra, (b) the X-ray light curves at 3.5~keV, 
and (c,d) hardness-intensity diagrams (HIDs). See Figs. \ref{Bchange} and \ref{Lchange}
for more explanations.}
\label{qchange}
\end{figure}

The effect of changes in the electron injection spectral index $q$ manifests itself 
obviously in the photon indices of the broadband spectra. Fig. \ref{qchange}a shows 
the time-averaged photon spectra for different values of $q$. We see that this 
spectral change results in a shift of the radiation peaks toward higher frequencies 
as the injection spectrum hardens. We see a decreasing value of the peak flux for 
higher values of $q$ (softer injection spectra) because of a decrease in the density 
of higher energy electrons with increasing injection spectral index. This causes a 
larger fraction of the injected energy to go into adiabatic rather than radiative 
losses (see Fig. \ref{cooling}a). Therefore, the jet becomes radiatively less 
efficient with increasing $q$. This also goes in tandem with a decreasing 
contribution from SSC, which follows directly from Eq. \ref{Lssc}.

As illustrated in the light curves in Fig. \ref{qchange}b, a higher value of $q$ leads 
to a steeper decline in flux once electron injection stops. At the energy of 3.55~keV,
the primary contribution is from the low-energy end of the Compton component, which
is increasing with increasing spectral index. For this reason, as a higher fraction
of the injected particle energy is injected near the low-energy cutoff of the electron 
distribution for increasing values of $q$, the peak medium-energy X-ray flux during 
the early phase of the simulated flares is positively correlated with $q$. During 
the later decay phase, the light curves converge to rather similar shapes and flux 
levels, though the decay slope is still positively correlated with the spectral 
index, as expected. 

The X-ray spectral hysteresis at energies just below and above the synchrotron cut-off 
is shifted according to the change in electron injection spectral index, but the overall 
characteristics remain unchanged (see Fig. \ref{qchange}c). The peak flux is reached at 
a higher value of the hardness ratio for a softer injection spectrum. The hardness ratio
decreases during the rising portion of the soft X-ray flux light curve, and increases 
during the decaying phase as the synchrotron component transits through the soft X-ray 
bands and gives way to an increasing contribution of the Compton components. The hard
X-ray flux is initially also dominated by synchrotron emission, except in the case of 
the softest injection index, $q = 2.8$. As the synchrotron component rapidly moves
out of the hard X-ray band, accompanied by a decreasing hardness ratio, the Compton
component becomes dominant, and a secondary, counterclockwise hysteresis loop develops
in the cases of soft electron injection spectra.

\subsection{\emph {Low-Energy Cutoff of the Electron Spectrum}} 

\begin{figure}[t]
\includegraphics[height=12cm]{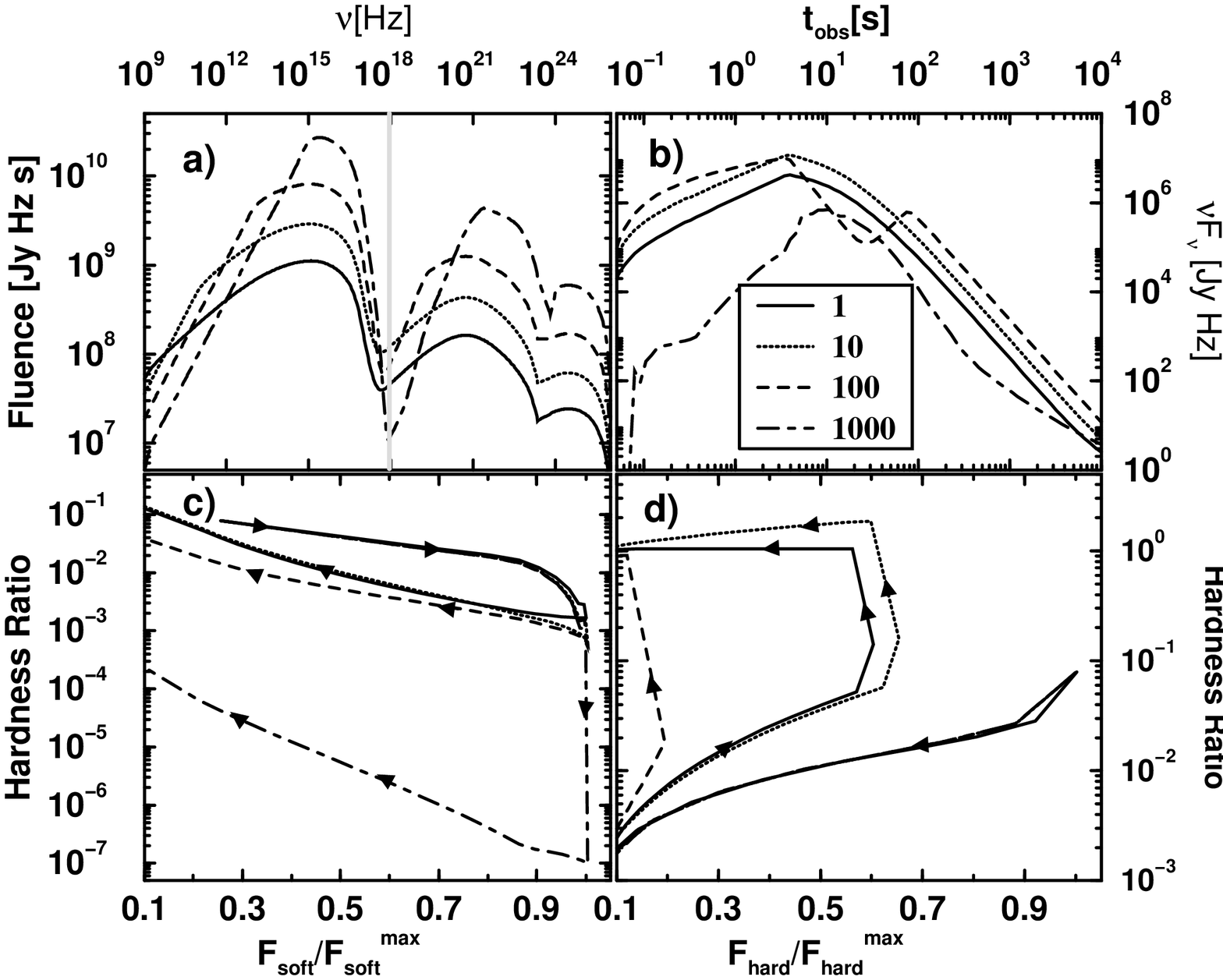}
\caption{Effect of a changing low-energy cutoff of the electron distribution, 
$\gamma_1$ (see legend), on (a) the time-integrated $\nu F_{\nu}$ (fluence) 
spectra, (b) the X-ray light curves at 3.5~keV, and (c,d) hardness-intensity 
diagrams (HIDs). See Figs. \ref{Bchange} and \ref{Lchange} for more explanations.}
\label{g1change}
\end{figure}

Since we have parameterized our injection spectra through an injection
luminosity and the bulk of the kinetic energy of electrons in the jet
is carried by the lowest-energy electrons for spectral indices of $q
> 2$, a change of the low-energy cutoff of the electron spectrum, $\gamma_1$,
naturally has quite dramatic effects on the radiative signatures from 
microquasar jets. These are illustrated in Fig. \ref{g1change}. Because 
of the dominance of adiabatic losses at low electron energies, the jet 
becomes radiatively more efficient for higher values of $\gamma_1$.
Consequently, the total radiative energy (fluence) increases, as also
seen in the previous section as a consequence of a harder electron 
injection spectrum. This also goes in tandem with an increased contribution
of SSC to the high-energy emission. A higher value of $\gamma_1$ shifts
the low-frequency cutoff of the individual radiation components towards
higher frequencies. However, since the $\nu F_{\nu}$ peak frequency is
dominated by emission from the highest-energy electrons for $q < 3$
(in the slow-cooling regime), a change of the value of $\gamma_1$ does
not affect the $\nu F_{\nu}$ peak frequency of the individual radiation
components. The shift of the high-energy peak into the $\gamma$-ray regime
for $\gamma_1 = 10^3$ as seen in Fig. \ref{g1change}a is a consequence of
the SSC component becoming dominant over the EC (star) component compared
to the lower values of $\gamma_1$.

The increasing contribution from SSC for $\gamma_1 = 10^2$ and $\gamma_1 = 10^3$
is also reflected in the light curves plotted in Fig. \ref{g1change}b. It is
evidenced in the faster onset, but then flatter rising portion of the flux
for $\gamma_1 = 10^2$ and the faster initial decay of that light curve, untill
the EC (star) component begins to move through the observing frequency (3.55~keV),
resulting in a similar decay as in the lower-$\gamma_1$ cases. In the case of
$\gamma_1 = 10^3$, the SSC component always dominates the 3.55~keV X-ray light 
curve.

The HIDs shown in Fig. \ref{g1change}c,d confirm these findings: In the
synchrotron-dominated portions of the X-ray flux (soft X-rays), the spectral 
hysteresis loops are clockwise and rather similar; however, for higher
values of $\gamma_1$, the hardness ratio becomes significantly lower because
of the increasing low-energy cutoff of the Compton spectra. The HIDs for
the hard X-ray flux show a similar trend as described in the previous section:
In particular, for low or moderate values of $\gamma_1$, a spectral softening
during the early flux decay is followed by a counterclockwise secondary 
hysteresis loop due to the influence of the EC (star) component.

\subsection{\emph {High-Energy Cutoff of the Electron Spectrum}} 

\begin{figure}[t]
\includegraphics[height=12cm]{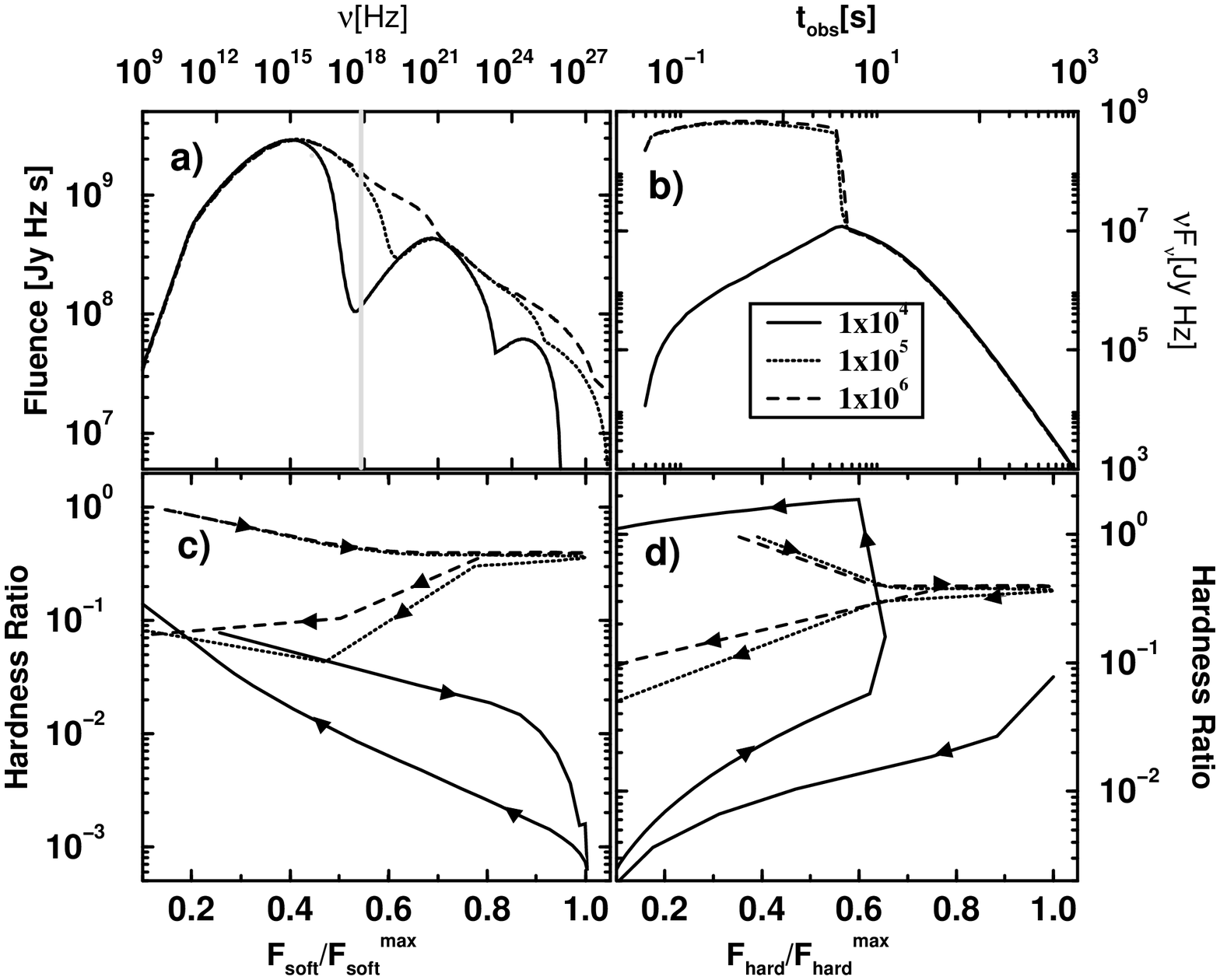}
\caption{Effect of a changing high-energy cutoff of the electron distribution, 
$\gamma_2$ (see legend) on (a) the time-integrated $\nu F_{\nu}$ (fluence) 
spectra, (b) the X-ray light curves at 3.5~keV, and (c,d) hardness-intensity 
diagrams (HIDs). See Figs. \ref{Bchange} and \ref{Lchange} for more explanations.}
\label{g2change}
\end{figure}

A higher cutoff of the electron injection spectrum will manifest itself
in the broadband spectra primarily by the extension of all radiation
components towards higher energies, as can be seen in Fig. \ref{g2change}a. 
Since only a minor portion of the overall particle energy is carried by
the highest-energy particles, the overall energetics of the particle
distribution and the bolometric luminosity and radiative energy output
of the emission region and the individual radiation components remain
virtually unchanged. For high values of $\gamma_2$, the individual
radiation components are increasingly overlapping, leading to a much
smoother overall shape of the broadband spectrum. 

The 3.55~keV light curves, plotted in Fig. \ref{g2change}b clearly show the
marked difference between the case of relatively low $\gamma_2$, where the
influence of the synchrotron component at 3.55~keV is very minor, and the
high-$\gamma_2$ cases, in which the synchrotron component extends significantly
beyond that energy. For our standard choice of $B_0 = 5$~kG and $D = 0.583$, 
the critical electron Lorentz factor for synchrotron emission at 3.55~keV 
is $\gamma_{\rm sy, 0} = 10^4$, while at the end 
of injection this value has increased to $\gamma_{\rm sy, 1} = 4 \times 
10^4$ due to the corresponding decline of the magnetic field over the 
injection length. The (co-moving) synchrotron cooling time scale for 
electrons emitting 3.55~keV synchrotron radiation at the end of the 
injection period is $t_{\rm sy} = 0.2$~s. Consequently, the synchrotron
contribution to the 3.55~keV light curve disappears virtually instantaneously
after the end of injection, so that the three light curves are basically
identical beyond that point. 

Fig. \ref{g2change}c reflects the fact that during the rising phase of the
light curve, the soft X-ray flux is clearly dominated by synchrotron emission.
In the high-$\gamma_2$ cases, the synchrotron cutoff is located beyond the
soft X-ray regime, leading to a relatively hard soft-X-ray spectrum that
gradually softens due to the development of a cooling break moving through 
the 0.1 -- 2~keV flux. For $\gamma_2 = 10^4$, the effect of radiative cooling
is immediately visible even in the soft X-ray regime, leading to a systematically
softer hardness ratio near the time of soft X-ray peak flux. The secondary
hard X-ray spectral hysteresis visible in the $\gamma_2 = 10^4$ case is not
evident in the higher-$\gamma_2$ cases just because of the much higher peak
flux, so that those hysteresis loops occur at much lower relative flux values
(compared to the maximum flux).

\subsection{\emph {Doppler Boosting Factor}}

\begin{figure}[t]
\includegraphics[height=12cm]{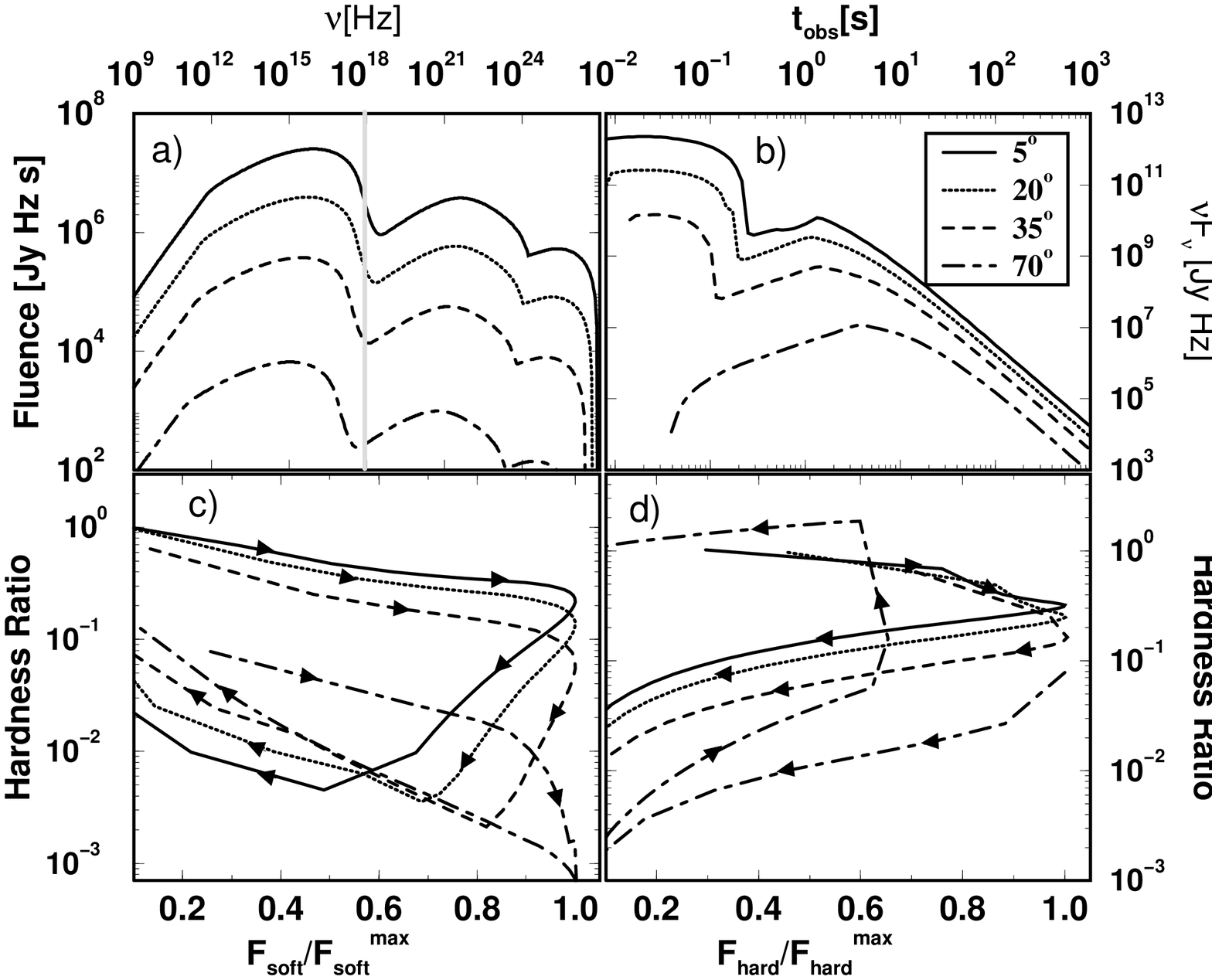}
\caption{Effect of a changing viewing angle (see legend), and thus, of the Doppler
boosting factor, on (a) the time-integrated $\nu F_{\nu}$ (fluence) 
spectra, (b) the X-ray light curves at 3.5~keV, and (c,d) hardness-intensity 
diagrams (HIDs). See Figs. \ref{Bchange} and \ref{Lchange} for more explanations.
The values of the Doppler factor $D$ corresponding to the various inclination angles 
are: $D (5^o) = 4.60$, $D (20^o) = 2.88$, $D (35^o) = 1.60$, $D (70^o) = 0.583$.}
\label{Dchange}
\end{figure}

We have investigated the effect of changing the Doppler factor $D$ on the 
spectra, light curves and HIDs by changing the observing angle $\theta_{obs}$. 
Fig. \ref{Dchange}a illustrates the spectral change resulting from a varying 
observing angle, given essentially by $\nu_{obs}\propto D$ and $\nu F_{\nu} 
\propto D^4$ for intrinsically isotropic emission \citep[note, however, that 
the dependence is somewhat stronger for EC emission, e.g.,][]{dermer95}. 

The 3.55~keV light curve (Fig. \ref{Dchange}b) and HIDs (Fig. \ref{Dchange}c,d)
illustrate the gradual shift of the synchrotron emission out of the X-ray regime
for lower Doppler factors (between individual runs), and a decaying magnetic field
as well as the development of the cooling break, moving through the X-ray energy
range (within individual runs). In particular, this leads to synchrotron-dominated
clockwise spectral hysteresis loops in the soft X-ray HIDs, and the emergence of
secondary, EC (star) dominated, counterclockwise hysteresis loops in the hard X-ray
HIDs for lower beaming factors. At times when the synchrotron component does no
longer have any influence on the X-ray light curves, the X-ray emission at 3.55~keV
is dominated by the EC (star) component, and we observe two separate regimes,
depending on the Doppler factor. For smaller Doppler boosting factors, the spectral
break corresponding to star light scattered by electrons that were injected at
$\gamma = \gamma_1$ occurs always below 3.55~keV. In that case, the peak in 
the X-ray light curve corresponds to the end of electron injection at $t_{\rm b, obs}
= t_1/D$. Consequently, it shows up at earlier observed times for increasing values 
of $D$. For higher values of the beaming factor, initially the observing frequency 
is below the break frequency corresponding to $\gamma_1$. In that case, the X-ray
light curve keeps rising even after the end of the injection period, and the peak 
corresponds to the time where the break frequency becomes lower than the observing 
frequency. Since this happens well after the end of electron injection, the time of 
peak flux in this case is in fact later than that for smaller beaming factors, which
explains the later light curve peak for $\theta = 5^o$ compared to $\theta = 20^o$.
Since the cooling of electrons injected at $\gamma_1 \lesssim 100$ is dominated by
adiabatic losses, their energy decreases as $\gamma(t) \propto t^{-m}$. Consequently,
we expect a dependence $t_{\rm b, obs} \propto D^{- 0.3/a}$ once the transition
to this regime has occurred.

\section{\label{summary}Summary and conclusions}

We have presented a detailed parameter study of the spectra
resulting from time-dependent 
injection and acceleration, and adiabatic and radiative cooling 
of nonthermal electrons in the jets of 
Galactic microquasars. Jet models of microquasars have recently attracted 
great interest, especially after the detection of VHE $\gamma$-ray emission 
from the high-mass X-ray binary and microquasar LS~5039, in combination with
the tentative identification of several microquasars with unidentified EGRET
sources. These detections have confirmed the idea that microquasars are 
a distinctive class of high and very high energy $\gamma$-ray sources.

It remains an open question whether the high-energy emission from 
microquasars is associated with leptonic or hadronic primaries. 
Predictions for the correlated variability at X-ray and 
high-energy $\gamma$-ray energies provides a discriminant
between the two possibilities. As shown here, X-ray hysteresis diagrams
are predicted in nonthermal leptonic models of microquasars, similar
to the situation for blazars. Detection of such variability behavior, 
particularly if correlated with $\gamma$-ray flaring behavior as 
described here, would provide evidence in favor of leptonic models. 

Whereas previous studies have largely focused on spectral fits to the broadband
emission of microquasars in a steady-state approximation, we have focused on the
spectral variability features expected in generic microquasar jet models with
electron injection and/or acceleration over a limited amount of time and length
along the jet, representative of, e.g., internal-shock models. We have conducted
a detailed parameter study, investigating the impact of variations of several
fundamental model parameters on the broadband SEDs, X-ray light curves, and the 
rapid X-ray spectral hysteresis phenomena expected to arise in these scenarios.
We have provided an analytical solution to the electron kinetic equation, taking 
into account radiation signatures of synchrotron, external Compton (with seed 
photons from the companion star and the accretion disk), as well as synchrotron 
self-Compton emission. In order to be able to work with an analytical solution
to the electron kinetic equation, we restricted our analysis to Compton scattering
in the Thomson regime. Consequently, our results are important to derive diagnostics
of microquasar jet emission in the X-ray regime, to contrast predictions of thermal
Comptonization models of these sources, but will be primarily applicable to 
microquasars that are not candidates of VHE $\gamma$-ray emission.
We have neglected the angle dependence of the stellar radiation field
\citep{boettcher05,db06}, which would lead to a modulation of the $\gamma\gamma$
absorption trough as well as the Compton scattered stellar radiation spectra depending on the 
orbital phase ($\phi_0=0$  showing the most absorption but highest Compton flux). The spectrum is 
expected to be further modified by pair cascades, redistributing some of the VHE radiative
power to lower frequencies, and significantly increasing the transparency of the source
\citep{ah05}. A study incorporating full Klein-Nishina effects on Compton scattering, 
the angle dependence of the stellar radiation field and cascade processes will be 
presented in a forthcoming publication on this subject.

Obviously, various spectral components (synchrotron, SSC, external-Compton) could
be easily distinguished if detailed snapshot SEDs could be measured for microquasars,
on the (often sub-second) time scales of their X-ray variability. Unfortunately,
such detailed snapshot broadband spectra are currently not available, and might 
not be available in the near future. Therefore, we have exposed several other
features pertinent to the transition between different spectral components which
will be more easily observable in realistic observational data of microquasars:

\begin{itemize}

\item A sudden increase of a light curve slope at a fixed observing frequency, 
not accompanied by significant flaring activity at other wavelengths, usually 
indicates the passing of a new spectral component through the fixed observing
frequency range. Most notably, this diagnostic can be used to investigate the
presence of one or more external-Compton component(s) in the X-ray / soft
$\gamma$-ray regime.

\item Clockwise spectral hysteresis in the hardness-intensity diagrams indicates
the dominance of synchrotron emission (in particular, before the end of the
injection period in our generic model setup). In this case, the frequency-dependent
light curve decay will be a useful diagnostic of the magnetic field strength in
the jet \citep[e.g.,][]{takahashi96}.

\item Counterclockwise spectral hysteresis in the hardness-intensity diagrams
indicates the dominance of Compton emission \citep[similar to the case of
blazars, see, e.g.,][]{bc02}. 

\item In our study, we found, quite often, a co-existence of clockwise and
counterclockwise X-ray hysteresis loops, which would provide
a particularly powerful diagnostic, as it would allow to probe the characteristic
transition energy between synchrotron and Compton emission, and its time
dependence. 

\end{itemize} 

There are only very few papers which present X-ray variability
in the form of hardness intensity diagrams. Those papers that do,
are presenting long-term variability information from RXTE
PCA observations, spanning time scales of hundreds of days, thus
representing the spectral characteristics of state transitions rather
than the short-term variability patterns that are the focus of our
analysis \citep[see, e.g.][]{fbg04,belloni05}. In such
a representation, possible hysteresis on short (intraday) time scales
are very hard to identify. An example where such short-term spectral
hysteresis may be identified, can be found in Homan et al. (2005: 
http://tahti.mit.edu/opensource/1655/), which shows an HID for GRO 1655-40. 
Here, within the overall long-term variability pattern, individual short 
flares (over a few days) can be discerned. One such outburst around MJD 53600 
shows a spectral hysteresis, which when transformed to the orientation of
the axes of HIDs in this paper, results in a counter-clockwise loop, 
indicating Compton dominated emission from the outflow for this outburst.

We conclude therefore that the X-ray variability as predicted by our model can be used 
as a diagnostic to gain insight into the nature of the high energy emission in microquasar 
jets. In particular, a transition between clockwise and counter-clockwise spectral
hysteresis would allow not only the distinction between different emission components,
but also parameters such as the magnetic field, the Doppler boosting factor, and the
characteristic electron injection / acceleration time.

\acknowledgments{The work of S.G. and M.B. was supported by NASA through XMM-Newton
GO grant no. NNG04GI50G and INTEGRAL theory grant NNG05GK59G. The work of C.D.D. is
supported by the Office of Naval Research and GLAST Science Investigation no.
DPR-S-1563-Y.}

\appendix

\section{Analytic Solution of the Electron Kinetic Equation}

The general electron cooling equation (\ref{cooling2}) can be solved
analytically by virtue of the substitution

\begin{equation}
y = \gamma \, t^m.
\end{equation}
which reduces Eq. (\ref{cooling2}) to
\begin{equation}
- {dy \over dt} = \left( \beta \, t^{-4a} + \delta t^{-2} + f_\ast (t)\right) \, t^{-m} \, y^2.
\label{reduced_general}
\end{equation}
This can easily be integrated to give
\begin{equation}
\begin{array}{rl}
{1 \over y} - {1 \over y_i} &= {\beta \over 1 - m - 4a} \left( t^{1 - m - 4a}
- t_i^{1 - m - 4a} \right)\cr  
&- {\delta \over 1 + m} \left( t^{-(1 + m)}
- t_i^{-(1 + m)} \right)\cr 
&+ \int_{t_i}^{t} f_{\ast} (t') \, {t'}^{-m} \, dt' \cr
\end{array}
\label{app_gen_y}
\end{equation}
Substituting $y = \gamma \, t^m$ yields the solution (\ref{general_g}) 
 in the main text.
 
The Jacobian in Eq. (\ref{N_gen_app}) is 
\begin{equation}
\left\vert {d\gamma_i \over d\gamma} \right\vert
\, \left\vert {dt_i \over d\gamma_i} \right\vert
= \left( {t_i \over t} \right)^m \, {\gamma^{-2} \over 
{m \over \gamma_i \, t_i} + {\beta \over t_i^{4a}} + {\delta \over t_i^{2}} + f_{\ast}(0)}
\label{jacobian_star}
\end{equation}
resulting in an electron spectrum at any given point along the jet:
\begin{equation}
N_e (\gamma, t) = {Q_0 \over \gamma^2} \int\limits_{\gimin}^{\gimax} 
d\gamma_i \; \left( {t_i \over t} \right)^m \; {\gamma_i^{-q} \over 
{m \over \gamma_i \, t_i} + {\beta \over t_i^{4a}} + {\delta \over t_i^{2}} + f_{\ast}(0)}.
\label{N_star_app}
\end{equation}

\end{document}